\documentclass[aps,prd,reprint,onecolumn,notitlepage,superscriptaddress]{revtex4-1}

\usepackage{lipsum}
\usepackage{soul}
\usepackage{amsmath}
\usepackage{graphicx}
\usepackage{color}
\usepackage{tikz}
\usepackage{verbatim}
\usepackage{hyperref}
\usepackage{amssymb}
\usepackage{bm}
\newcommand{\rom}[1]{\textup{\uppercase\expandafter{\romannumeral#1}}}

\allowdisplaybreaks

\begin{document}

\title{Local momentum space: Scalar field and gravity}

\author{Sukanta Panda \footnote{email:sukanta@iiserb.ac.in}
       }

\affiliation{Indian Institute of Science Education and Research Bhopal,\\ Bhopal 462066, India}

\author{Abbas Tinwala \footnote{email:abbas18@iiserb.ac.in}
       }

\affiliation{Indian Institute of Science Education and Research Bhopal,\\ Bhopal 462066, India}

\author{Archit Vidyarthi \footnote{email:archit17@iiserb.ac.in}
       }

\affiliation{Indian Institute of Science Education and Research Bhopal,\\ Bhopal 462066, India}

\begin{abstract}
We use the local momentum space technique to obtain an expansion of the Feynman propagators for scalar field and graviton up to first order in the background curvature. The expressions for the propagators are cross-checked with the past literature as well as with the expressions for the traced heat kernel coefficients. The propagators so obtained are used to compute one-loop divergences in the Vilkovisky-Dewitt's effective action for a scalar field non-minimally coupled with gravity for an arbitrary spacetime metric background. The Vilkovisky-DeWitt effective action is then compared with the standard effective action in the limit $\kappa =0$, where $\kappa = 2/M_P$ in terms of the Planck mass. The comparison yields the important result that taking the limit $\kappa=0$ after computing the Vikovisky-DeWitt effective action is not equivalent to computing the Vikovisky-DeWitt effective action for the same theory in the absence of gravity.
\end{abstract}



\maketitle

\section{Introduction}
The standard effective action in gauge theories suffers from the problem of off-shell gauge dependence and dependence on the choice of field parametrization. A substantial number of evidence exists to lend credence to the statement made above regarding off-shell gauge dependence as follows. In \cite{robinson}, authors performed a calculation using standard effective formalism in quantized Yang-Mills-Einstein gravity and discovered quantum gravity contribution to the $\beta$ function for gauge coupling parameter in the Yang-Mills theory. The remarkable outcome of this calculation was that the gauge theories that were not asymptotically free in the absence of gravity became asymptotically free when the quantum gravity contribution to the $\beta$ function was taken into account. However, the calculations in \cite{robinson} were shown to be ambiguous in \cite{artur}, where the authors performed a similar calculation in a different gauge and surprisingly found no quantum gravitational contribution to the $\beta$ function. In another paper \cite{kaluza1}, the authors demonstrated that quantum effects could explain the smallness of extra dimensions by computing the effective potential using the standard effective action theory. The result of this paper was also found in \cite{kaluza2,kaluza3} to suffer from the problem of gauge dependence. Some other works that highlight the issue of gauge dependence in the standard effective action are \cite{gd1,gd2}.

   To resolve this issue the standard effective action formalism was refined by Vilkovisky and further modified by DeWitt. The new formalism gives an effective action, known as the Vilkovisky-DeWitt(VD) effective action, which by construction is completely gauge-invariant, gauge condition independent, and field parametrization independent (for details see \cite{unique,tomsbook, Odintsovbook}). The new formalism raises an important question: ``Under what conditions do the VD effective action and the standard effective action match?". An obvious one is when the background fields are chosen to satisfy the classical equations of motion (see \cite{ourpaper1}). Another case in which the two effective actions must agree upon is when they are computed off-shell for certain non-gauge theories. While making this statement, we must exercise caution. This is because the standard effective action also suffers from dependence on the choice of field parametrization (in fact, this is the very cause of gauge dependence in the off-shell standard effective action). Therefore for non-gauge theories involving multiple fields like non-linear $\sigma$ theory (see chapter-6 of \cite{tomsbook}), it still becomes mandatory to use VD formalism to obtain a parametrization independent effective action since there is no preferred coordinate system in which the fields can take values.

   In \cite{mackaytoms} it was shown that the VD effective action, for the theory of a scalar field coupled to gravity, was identical to the standard effective action in the limit $\kappa = 0$, where $\kappa = 2/M_P$ in terms of the Planck mass. There, the authors chose a flat Euclidean spacetime metric as the background. The purpose of taking the limit $\kappa =0$ was to turn off gravity. Does this mean that taking $\kappa=0$ after computing the VD effective action is equivalent to computing the VD effective for the theory of a scalar field where the spacetime metric is only a background field? If so, it makes sense for the effective actions to match since dropping the Einstein-Hilbert term $R/\kappa^2$ from the classical action makes it a non-gauge theory, for which the effective actions are expected to match off-shell.

   We wish to see whether the effective actions match when $\kappa$ is taken zero at the end of the calculation for the same theory by choosing an arbitrary metric background unlike the flat Euclidean as chosen in \cite{mackaytoms}. By doing so, we would like to check whether taking $\kappa =0$ after computing the effective action is equivalent to starting with a non-gauge theory and then computing the effective action.
  
  The classical action of the theory of our interest in the Jordan frame is,
\begin{align}
    S = \int d^4x\sqrt{|g(x)|}\left(-\frac{2R}{\kappa^2} + \frac{1}{2}g^{\mu\nu}\nabla_\mu\phi\nabla_\nu\phi + \frac{1}{2}(m^2+\xi R)\phi^2 + \frac{1}{24}\lambda\phi^4\right).\label{action}
\end{align}

   We will use the technique of local momentum space to compute the one-loop divergences in the VD effective action in the Jordan frame for an arbitrary metric background. Before we lay a proper outline of this work a brief description of the technique of local momentum space is in order.
    
   The technique of local momentum space was originally introduced by Bunch and Parker \cite{bunch} to study the renormalization of UV divergences of interacting scalar and spin 1/2 fields in curved space-time. The technique is based on the principle of equivalence in which a curved space-time can be seen to be locally flat. As a result, local momentum space representation makes it possible to apply standard momentum space techniques to curved spacetimes. The basic formalism consists of constructing Riemann normal coordinates with the origin fixed at point $x'$ and the variable space-time point $x$ being in the neighborhood of $x'$. This enabled the authors in \cite{bunch} to obtain a local momentum space expansion (Fourier transform with respect to $x$) of the Feynman propagators for scalar and spin 1/2 fields. In a subsequent work by Bunch \cite{bunch2} the technique was generalized by obtaining a momentum space representation for the scalar field propagator $G(x,x')$ in which both $x$ and $x'$ may be variable with a third point $z$ chosen to be fixed. The purpose of this was to study the two-loop renormalizability of $\lambda\bar\phi^4$ theory in curved space-time. Since then, many other applications have emerged from this technique over the years. There are several earlier examples, including the demonstration of curvature-induced asymptotic freedom in \cite{curvasymp}, the application to the Kaluza-Klein theory in \cite{kaluza4,kaluza5,kaluza6}, and some other applications in \cite{otherapp1,otherapp2,otherapp3,otherapp4}. A few years back, the technique was used to examine how quantum gravitational corrections affect gauge coupling constants \cite{tomsapp1,tomsapp2}. The work in \cite{tomsapp1} is particularly noteworthy as it shows that the quantum gravitational effects nullify electric charge at high energies as opposed to what happens when non-gravitational effects alone are taken into account. Most recently the technique has been used to evaluate divergences in effective action for gauged Yukawa model in curved space-time. \cite{tomsapp4}. Another application was seen in \cite{tomsapp3} where the authors introduce methods to obtain heat kernel coefficients for non-minimal operators. One additional purpose besides the ones mentioned earlier is to correct the results obtained for the heat kernel coefficients in \cite{tomsapp3}. 
    
The outline of the paper is as follows. In \hyperref[SecII]{Sec.II} we prepare the necessary ingredients which are prerequisites for the one-loop VD effective action computation. In \hyperref[SecIII]{Sec.III}, we use the local momentum space technique to obtain the expressions for the scalar and graviton propagators (up to the first order in the background curvature). This is followed by a few necessary cross-checks with some remarks on previous work. In \hyperref[SecIV]{Sec.IV}, we write down explicitly the expressions for the divergences in one-loop VD and standard effective actions followed by a couple of cross-checks. Finally, in \hyperref[SecV]{SecV} we perform renormalization with a detailed discussion on how the effective actions compare in the limit $\kappa=0$. The calculations being weighty throughout the paper we utilize Mathematica. 

Before we end this section let us familiarise ourselves with the DeWitt notation, which we will be following in this paper. The symbols used to represent the properties of a field are labeled as $\varphi^i$, where the discrete field index and the field's space-time argument are condensed into the single label $i$. For instance, if the field is a scalar field, then $\varphi^i$ is equivalent to $\phi(x)$, and if it is a vector field, then $\varphi^i$ is equivalent to $A_\mu(x)$, while for a second-rank tensor field, $\varphi^i$ is equivalent to $g_{\mu\nu}(x)$. Additionally, the following summation convention is used in $n$ dimensions:

\begin{equation}\label{2}
    \varphi^i B_{ij} \varphi^j = \int d^nx\int d^nx' \varphi^I(x) B_{IJ}(x,x') \varphi^J(x'),
\end{equation}

\noindent
where capital Latin letters $(I, J,...)$ are used as a placeholder for conventional field indices,

\noindent
Also from,

\begin{align}\label{3}
     \varphi^I(x) &= \int d^nx' |g(x')|^{1/2}\delta(x,x') \varphi^I(x'),\nonumber\\
     & = \int d^nx' \tilde\delta(x,x') \varphi^I(x'),
\end{align}

\noindent
 we define $\tilde\delta(x,x')$:
 
\begin{equation}\label{4}
    \tilde\delta(x,x') = |g(x')|^{1/2}\delta(x,x'),
\end{equation}

\noindent
where $\delta(x,x')$ is the conventional bi-scalar Dirac $\delta$-distribution.

We also choose the $(-,+,+,+)$ convention for the flat Minkowski metric wherever used throughout the paper.
    
    \section{Some essential preliminaries for the computation of divergences in the one-loop VD effective action}\label{SecII}
    
In the introduction, we mentioned that our main purpose is to compare the VD effective action with the standard one, for which a theory of scalar field and gravity is a simple and adequate choice. In \cite{mackaytoms} it was shown that the effective actions match up to the second order in the background scalar field in the limit $\kappa = 0$. Since the computations there, were performed for a flat Euclidean metric background, we would like to see how the effective actions compare up to the next immediate order in the background curvature. In this light, we shall compute the VD effective action only up to the first order in background curvature and the second order in the background scalar field. We would like to point out that even up to the mentioned order, computing the divergences in the VD effective action by employing the technique of local momentum space is barely manageable. 

Putting aside the intricacies of Vilkovisky-DeWitt's procedure (the details of which can be found in \cite{dewitt, unique, tomsbook, Odintsovbook}) for obtaining gauge-invariant and gauge condition independent effective action, we only state the expression to one-loop order here,
   \begin{equation}\label{effac}
        \Gamma[\bar\varphi] = -\text{ln det} Q_{\alpha\beta}[\bar\varphi] + \dfrac{1}{2}\lim_{\alpha\rightarrow 0}\text{ln det}\Big(S^{;i}{}_j + \dfrac{1}{2\alpha}K^i_\beta[\bar\varphi]K^\beta_j[\bar\varphi]\Big),
    \end{equation}
    where $Q_{\alpha\beta}$ is the ghost term, $K^i_\beta$ are the generators of gauge transformation and $S$ is the classical action. The gauge parameter $\alpha$ (not to be confused with the spacetime index) must be taken zero after the effective action has been calculated. The main difference between this expression and that of standard effective action lies in the use of the covariant derivative,
    \begin{align}
        S_{;ij} = S_{,ij} - \Gamma^k_{\ ij}S_{,k},
    \end{align}
    where $\Gamma^k_{\ ij}$ are the connections that are indispensable for obtaining a gauge condition-independent result.
     There is more than one way to compute the above expression. One way is to find the differential operator corresponding to $S^{;i}{}_j + \dfrac{1}{2\alpha}K^i_\beta[\bar\varphi]K^\beta_j[\bar\varphi]$ and then employ the heat kernel technique (see chapters 5-7 of \cite{tomsbook}, and \cite{scholar}) to compute the divergences in effective action. This technique applies only to some restricted class of operators called minimal operators. Although the technique can be extended to non-minimal operators as well following the work of Barvinsky and Vilkovisky \cite{nonminimal} (see \cite{ourpaper1,od1,od2,od3,od4,od5,od6} for its applications), it is still very difficult to implement it in practice. As such we resort to the perturbative expansion technique as follows (for application of this technique see  \cite{ourpaper2,mackaytoms,tomsapp4,sandeep}). We first split the fields $\varphi^i$ into a background $\bar\varphi^i$ and a "quantum part" $\eta^i$ as follows,
     \begin{equation}\label{split1}
        \varphi^i = \bar\varphi^i + \eta^i.
    \end{equation}

 Remember that $\varphi^i$ could either be the scalar field or the metric field depending on the space-time indices contained in the condensed index `$i$' on the field $\varphi^i$. We can now write the term in brackets in Eq.~(\ref{effac}) as,
    \begin{equation}\label{pert}
        -\dfrac{1}{2}\text{ln det}\Big(S^{;i}{}_j + \dfrac{1}{2\alpha}K^i_\beta[\bar\varphi]K^\beta_j[\bar\varphi]\Big) = \text{ln}\int [d\eta]\exp\{-S_\text{q}\},
    \end{equation}
    where $S_{\text{q}}$ is given by,
    \begin{align}\label{Sq}
    &S_{\text{q}} = \dfrac{1}{2}\eta^i(S_{,ij} - \Gamma^k_{ij}S_{,k})\eta^j + S_\text{gb},
        \end{align}
    and $S_{\text{gb}}$ is the gauge-breaking action given by,    
        \begin{align}\label{Sgb}
        &S_\text{gb} = \dfrac{1}{4\alpha}\eta^i K^\beta_i K_{j\beta}\eta^j.
        \end{align}
    
    The derivatives wrt. the fields of $S$ appearing in Eqs.~\ref{Sq}, for instance, $S_{,ij}$, are to be computed at the background value of the fields.
    Following the expansion of $S$ in powers of the background field up to the second order,
    \begin{align}
        S_\text{q}=S_0+S_1+S_2,
    \end{align}
    where the subscript attached to $S$ indicates the power of the background field it consists of,
    we can show that,
     \begin{align}
         \text{ln}\int [d\psi]\exp\{-S_\text{q}\} &= -\langle S_1\rangle - \langle S_2\rangle + \dfrac{\langle (S_1)^2\rangle}{2} + \mathcal{O}(\bar\varphi^3)
     \end{align}
     The angular brackets represent the expectation value in the path integral formulation. For instance,
     
     \begin{align}
         \langle S_1\rangle = \dfrac{\int [d\psi]e^{-S_0}S_1}{\int [d\psi]e^{-S_0}}
     \end{align}
     
    We finally obtain the one-loop effective action up to quadratic order in the background field,
     \begin{equation}\label{finaleffac}
     \Gamma[\bar\varphi] = -\text{ln det} Q_{\alpha\beta} + \lim_{\alpha\rightarrow 0}\Big(\langle S_1\rangle + \langle S_2\rangle - \dfrac{\langle (S_1)^2\rangle}{2}\Big).
     \end{equation}
Let us now get a few ingredients ready that are necessary to compute the above expression. We start with the generators of gauge transformation:
\begin{align}
    &K^1_{\mu\nu \ \alpha}(x,x') = -(\partial_\alpha g_{\mu\nu}(x) + g_{\mu\alpha}(x)\partial_\nu + g_{\nu\alpha}(x)\partial_\mu)\tilde\delta(x,x'),\nonumber\\
    &K^2_\alpha(x,x') = -\partial_\alpha\phi(x)\tilde\delta(x,x'),\label{gen}
\end{align}
where the index represents the fields:  $g_{\alpha\beta} \rightarrow 1 ,\hspace{3mm} \phi\rightarrow 2$. We use this additional convention when we need to spell out explicitly the ``field" components of a quantity. For instance, here the generator of gauge transformation has two components.

 When explicitly written, Eq.~\ref{split1} would read
    \begin{align}
        &g_{\mu\nu} = \bar g_{\mu\nu}+\kappa h_{\mu\nu},\nonumber\\
        &\phi = \bar\phi + \psi,\label{split}
    \end{align}
    where we have declared the fields carrying a bar over them as the background field and the other piece as the quantum field.

The field space metric has four components. It takes a non-diagonal form owing to the presence of the derivative coupling between the scalar field and the metric coming from the non-minimal coupling (see \cite{ourpaper1}) (like $\partial_\mu\phi\partial_\nu g_{\alpha\beta}$),
\begin{align}
   &G_{11}^{\alpha\beta\mu\nu}(x,x') = \frac{\sqrt{\bar g(x)}}{2}F(\bar\phi)(2\bar g^{\alpha(\mu}(x)\bar g^{\nu)\beta}(x) -\bar g^{\alpha\beta}(x)\bar g^{\mu\nu}(x))\tilde\delta(x,x'),\nonumber\\
   &G_{12}^{\alpha\beta}(x,x')=\sqrt{\bar g(x)}H(\bar\phi) \bar g^{\alpha\beta}(x)\tilde\delta(x,x'),\nonumber\\
   &G_{22}(x,x') = \sqrt{\bar g(x)}J(\bar\phi)\tilde\delta(x,x'), \label{metric}
\end{align}
where the brackets in the indices represent symmetrization in the pair ($\mu$,$\nu$). The functions appearing in the expressions above read,
\begin{align}
&F(\bar\phi)=1-\frac{\kappa^2\xi\bar\phi^2}{4},\nonumber\\
&H(\bar\phi)=\frac{\kappa\xi\bar\phi}{2},\nonumber\\
&J(\bar\phi)=1+\frac{\kappa^2\xi^2\bar\phi^2}{2}. \label{metricfunctions}
\end{align}
Let us note that the choice of the field space metric has been fixed with the prescription that the field space metric is determined from the highest derivative term in the bilinear form of the action in the minimal gauge, i.e., $\alpha =1$. Such a choice has been justified in \cite{ourpaper1} where the authors also point out why the choice of the field space metric in \cite{mackaytoms} is incorrect. We suggest \cite{shapiro} for a detailed discussion on the choice of field space metric. 
 
 Let us now compute the VD connections $\tilde\Gamma^i_{jk}$. The choice of gauge matters here as it can lead to considerable simplification in the expression of connections. In particular, it can be shown (see \cite{fredkin}, or, for a pedagogical treatment see chapter 7 of \cite{tomsbook}) that if we choose the Landau-DeWitt gauge,
 \begin{align}
   \chi_\alpha =  K^i_\alpha[\bar\varphi]g_{ij}[\bar\varphi](\varphi^j-\bar\varphi^j)=0,\label{ldg}
    \end{align}
   then the complicated VD connections $\tilde\Gamma^k_{\ ij}$ can be replaced by simple Christoffel connections $\Gamma^i_{jk}$ over field space. These have been written down in Appendix \ref{AppA} along with the inverse field space metric. Upon using Eqs. (\ref{gen}), (\ref{metric}), (\ref{metricfunctions}) and (\ref{split}) in (\ref{ldg}) we find,
    \begin{align}
        \chi_\alpha = \frac{2}{\sqrt{F(\bar\phi)}}\sqrt{|\bar g(x)|}\Big(\frac{\nabla^\mu (F(\bar\phi)h_{\mu\alpha})}{\kappa} - \frac{\bar g^{\mu\nu}\nabla_\alpha(F(\bar\phi) h_{\mu\nu})}{2\kappa}&+\frac{\nabla_\alpha(\psi H(\bar\phi))}{\kappa}-\dfrac{1}{2}hH(\bar\phi)\nabla_\alpha\bar\phi-\dfrac{1}{2}\psi J(\bar\phi)\nabla_\alpha\bar\phi\Big),
    \end{align}
    where we have inserted a factor of $\sqrt{F(\bar\phi)}$ which does not alter the gauge. The gauge-breaking Lagrangian following the gauge condition reads,
    \begin{align}
        S_{\text{gb}} = \frac{\kappa^2}{F(\bar\phi)\alpha}\int d^4x \sqrt{|\bar g(x)|}\Big(\nabla^\mu (F(\bar\phi)h_{\mu\alpha}) - \frac{1}{2}g^{\mu\nu}&\nabla_\alpha(F(\bar\phi) h_{\mu\nu})-\dfrac{1}{2}hH(\bar\phi)\nabla_\alpha\bar\phi\nonumber\\&-\dfrac{1}{2}\psi J(\bar\phi)\nabla_\alpha\bar\phi+\nabla_\alpha(\psi H(\bar\phi))\Big)^2.
    \end{align}
    We now proceed to calculate $S_0$, $S_1$, and $S_2$ as outlined earlier. We obtain
    \begin{align}
       &S_0 =  \int d^4x\sqrt{|\bar g(x)|}\Bigg[\frac{1}{2}m^2\psi^2+\left(\frac{\omega}{4}+\dfrac{\xi}{2}(1-\omega)\right)R\psi^2+\frac{1}{2}\nabla_\mu\psi\nabla^\mu\psi+\frac{1}{2}\nabla_\alpha h_{\mu\nu}\nabla^\alpha h^{\mu\nu}\nonumber\\&\hspace{30mm} - \nabla_\beta h_{\alpha\mu}\nabla^\mu h^{\alpha\beta}+\frac{1}{\alpha}\nabla_\beta h^{\alpha\beta}\nabla^\mu h_{\alpha\mu} + \left(1-\frac{1}{\alpha}\right)\nabla_\mu h \nabla_\alpha h^{\mu\alpha} \nonumber\\&\hspace{30mm}+ \frac{1}{2}\left(\frac{1}{2\alpha} - 1\right)\nabla_\mu h\nabla^\mu h
        +(\omega-2) h^{\alpha\mu}h_\mu^{ \ \beta} R_{\alpha\beta} + \left(1-\frac{\omega}{2}\right) h h^{\mu\nu}R_{\mu\nu} \nonumber\\&\hspace{30mm}+ \frac{1}{2}\left(1-\frac{\omega}{2}\right)h^{\mu\nu}h_{\mu\nu}R - \frac{1}{4}\left(1-\frac{\omega}{2}\right)h^2R\Bigg],\label{s0}\\
       &S_1 = \kappa\int d^4x \sqrt{|\bar g(x)|}\Bigg[M_1^{\mu\nu}(\bar\phi)\psi h_{\mu\nu}+M_2^{\mu\nu\alpha}(\bar\phi)h_{\mu\nu}\nabla_\alpha\psi+M_3^{\mu\nu\alpha\beta}(\bar\phi)h_{\mu\nu}\nabla_\alpha\nabla_\beta\psi\Bigg],\label{s1}\\
       &S_2 = \kappa^2\int d^4x \sqrt{|\bar g(x)|}\Big[N_1(\bar\phi)\psi^2+N_2^{\mu}(\bar\phi)\psi\nabla_\mu\psi+N_3^{\mu\nu}(\bar\phi)\psi\nabla_\mu\nabla_\nu\psi+N_4^{\mu\nu\alpha\beta}(\bar\phi)h_{\mu\nu}h_{\alpha\beta}\nonumber\\&\hspace{30mm}+N_5^{\mu\nu\alpha\beta\gamma}(\bar\phi)h_{\mu\nu}\nabla_{\gamma}h_{\alpha\beta}+N_6^{\mu\nu\alpha\beta\gamma\tau}(\bar\phi)h_{\mu\nu}\nabla_\tau\nabla_\gamma h_{\alpha\beta}\Big].\label{s2}
       \end{align}
       The expression of various functions appearing in $S_1$ and $S_2$ have been written down in Appendix \ref{AppB}.
       We have introduced a parameter $\omega$ which appears as a factor accompanying the field space connections. Mathematically we have just made a replacement as follows.
       \begin{align}\label{omega}
       \Gamma^i_{jk}\rightarrow \omega\Gamma^i_{jk}.
       \end{align}
       The purpose of introducing $\omega$ is to keep track of terms arising from connections. In this way, we can turn off the terms arising from connections by taking $\omega=0$, which is necessary to derive the standard effective action from the VD effective action.
       Having furnished all the necessary ingredients we are all set now to proceed for the local momentum space expansion to get the propagators from $S_0$ in the following section.
       \section{Local momentum space expansion}\label{SecIII}
       For the computation of effective action, we need the propagators which are determined by $S_0$. The equations satisfied by Green's function can be extracted from $S_0$. The Green's function for the scalar field satisfies,
       \begin{align}
           \sqrt{|g(x)|}\Big(-g^{\alpha\beta}\nabla_\alpha\nabla_\beta + m^2 + \left(\dfrac{\omega}{2}+\xi(1-\omega)\right)R\Big)G(x,y) = \delta(x,y),\label{seq}
       \end{align}
       and the Green's function for the graviton satisfies,
       \begin{align}
           \sqrt{|g(x)|}\Big(A^{\alpha\beta\mu\nu\rho\sigma}\nabla_\rho\nabla_\sigma + B^{\alpha\beta\mu\nu}\Big)G_{\mu\nu\gamma\tau}(x,y) = \delta(x,y)\delta^{\alpha\beta}_{\gamma\tau},\label{geq}
       \end{align}
       where
       \begin{align}
           &A^{\alpha\beta\mu\nu\sigma\rho} = \left(1-\frac{1}{2\alpha}\right)g^{\alpha\beta}g^{\mu\nu}g^{\rho\sigma} - g^{\alpha(\mu}g^{\nu)\beta}g^{\rho\sigma} -\frac{2}{\alpha}g^{\rho(\alpha}g^{\beta)(\mu}g^{\nu)\sigma} + 2g^{\sigma(\alpha}g^{\beta)(\mu}g^{\nu)\sigma}\nonumber\\
           &\hspace{15mm}+ \left(\frac{1}{\alpha} -1\right)(g^{\alpha\beta}g^{\rho(\mu}g^{\nu)\sigma} + g^{\mu\nu}g^{\rho(\alpha}g^{\beta)\sigma}),\\
           &B^{\alpha\beta\mu\nu} = (\omega-2)g^{\alpha(\mu}R^{\nu)\beta}-(\omega-2)g^{\beta(\mu}R^{\nu)\alpha} + \left(1-\frac{\omega}{2}\right)g^{\alpha\beta}R^{\mu\nu}+\left(1-\frac{\omega}{2}\right)g^{\mu\nu}R^{\alpha\beta} \nonumber\\
           &\hspace{10mm}+ \left(\frac{1}{2}-\frac{\omega}{4}\right)(2g^{\alpha(\mu}g^{\nu)\beta} - g^{\alpha\beta}g^{\mu\nu})R,\\
           &\delta^{\alpha\beta}_{\mu\nu} = \frac{1}{2}(\delta^\alpha_\mu\delta^\beta_\nu + \delta^\alpha_\nu\delta^\beta_\mu).
       \end{align}
       The focus of this section is to solve equations (\ref{seq}) and (\ref{geq}) to obtain Green's functions. The technique that we will employ for this purpose is that of Bunch and Parker \cite{bunch}. We introduce the Riemann normal coordinates $y^\mu$ and $x^\mu$ of point $y$ and $x$ respectively taking point $z$ as the origin. This is what is known as the Riemann normal coordinate expansion about a third point first employed in \cite{bunch2}. The reason to expand about a third point here is because of the appearance of the derivatives of quantum fields in expressions for $S_1$ and $S_2$. These can be recast as derivatives of propagators that diverge in the limit $x \rightarrow y$. In the absence of derivatives, these could have been dealt with easily by expanding about $y$. However, to find divergences in the derivatives of propagators we first need to find the derivatives and then take the limit $x \rightarrow y$ making it necessary for us to expand the propagators about a third point. The standard Riemann normal coordinate expansions at point $y$ up to the first order in curvature that will be necessary for our work are as follows.
       \begin{align}
           &g_{\mu\nu}(y) = \eta_{\mu\nu} - \frac{1}{3}y^\alpha y^\beta R_{\mu\alpha\nu\beta}|_{y=0}+\mathcal{O}(R^2),\nonumber\\
           &g^{\mu\nu}(y) = \eta^{\mu\nu} + \frac{1}{3}y^\alpha y^\beta R^{\mu\ \nu}_{\ \alpha\ \beta}|_{y=0}+\mathcal{O}(R^2),\nonumber\\
           &\Gamma^\mu_{\ \rho\sigma}(y) = -\frac{2}{3}y^\alpha R^\mu_{\ (\rho\sigma)\alpha}|_{y=0}+\mathcal{O}(R^2),\nonumber\\
           &|g(y)| = 1 - \frac{1}{3}y^\alpha y^\beta R_{\alpha\beta}|_{y=0}+\mathcal{O}(R^2).\label{srnc}
       \end{align}
       The curvature tensors that appear in the above expansions are basically the coefficients of the Taylor expansion which must be evaluated at the origin, i.e., at $y = 0$. The raising and lowering of indices are done with the Minkowski metric once an expansion in Riemann normal coordinates is performed.
       Following normal coordinate expansion we make use of Fourier transformation to write
       \begin{align}
           &G(x,y) = \int \frac{d^4k}{(2\pi)^4}e^{iku}\tilde G(k,z),\nonumber\\
           &G_{\mu\nu\gamma\tau}(x,y) = \int \frac{d^4k}{(2\pi)^4}e^{iku}\tilde G_{\mu\nu\gamma\tau}(k,z),\label{ftp}
       \end{align}
       where $u^\mu=x^\mu-y^\mu$. The Fourier-transformed Green's functions depend upon the origin $z$ but henceforth we will not specify this explicitly. We then assume that the momentum space Green's functions can be expanded out asymptotically in the inverse powers of $k$ as follows
       \begin{align}
           \tilde G(k) =  \tilde G_0(k)+ \tilde G_1(k)+ \tilde G_2(k) + ... ,\label{pexp}
       \end{align}
       with a similar expression for the graviton propagator.
       The subscript counts the order of the derivative of the metric, i.e., $\tilde G_0$ will not contain any derivative of metric, $\tilde G_1$ will contain first order derivative of metric, $\tilde G_2$ will contain second order derivative of metric and so on. This is obvious because technically, Riemann normal coordinate expansion makes use of local flatness and so the coefficients would involve derivatives of the metric. To balance the successive order of derivatives we require momentum factors in the denominator which justifies the earlier assumption. Because the propagators are expanded in inverse powers of $k$, the whole technique is applicable only for large momentum. As a result, we will completely ignore any divergences that arise from $k\rightarrow 0$ in the effective action.
       
       The procedure is to use Eq.~(\ref{srnc}), (\ref{ftp}) and (\ref{pexp}) to expand the equations (\ref{seq}) and (\ref{geq}) in orders of the derivatives of the metric. The expressions that will follow will involve factors of $u^\mu$ which can be dealt with by using,
       \begin{align}
           u^{\mu_1}u^{\mu_2}...u^{\mu_n}e^{iku} = \frac{1}{i^n}\frac{\partial^n}{\partial k_{\mu_1}\partial k_{\mu_2}...\partial k_{\mu_n}}e^{iku}.
       \end{align}
       The derivatives acting on exponential can then be discarded making use of partial integrations with respect to k.
       \subsection{Propagators}\label{SecIIIa}
       Following the procedure outlined earlier we find the following result for the scalar field propagator up to first order in curvature\cite{bunch2},     
    \begin{align}
        &\langle \psi(x)\psi(x')\rangle = \int \frac{d^4k}{(2\pi)^4}e^{iky}\tilde G(k),\nonumber\\
        &\tilde G(k) = \frac{1}{k^2+m^2}+\left(\frac{1}{3}-\frac{\omega}{2}+\sigma(1-\omega)\right)\frac{R}{(m^2+k^2)^2} - \frac{2R_{\alpha\beta}k^\alpha k^\beta}{3(m^2+k^2)^3}\nonumber\\&\hspace{75mm} +\dfrac{R_{\alpha\beta}y^\alpha y^\beta}{6(m^2+k^2)}-\dfrac{R_{\alpha\mu\beta\nu}k^\alpha k^\beta y^\mu y^\nu}{3(m^2+k^2)^2}.\label{sprop}
    \end{align}
    Apart from the connection-dependent terms (proportional to $\omega$), the expression for the scalar field propagator is identical to that obtained in \cite{bunch2}.
    
 The full graviton propagator up to first order in curvature is too lengthy to be written here so we have mentioned it in the Appendix \ref{AppC}. Some important cross-checks of the graviton propagator from previous works are in order. The expression for the graviton propagator derived in earlier works involved taking $y$ as the origin. In that case, if we turn off the connection terms in Eq.(\ref{G2g})  by taking $\omega = 0$ and set the gauge parameter $\alpha$ to 1 then the expression coincides with that found in \cite{odintsovgprop}. Since the gauge parameter $\alpha$ has been kept unspecified, the expression for the graviton propagator presented in our work is, therefore, more general and most importantly can be taken to be applicable for the computation of the VD effective action since we must take $\alpha\rightarrow 0$ (see Eq. \ref{effac} and the lines following it) after having computed the effective action (\ref{effac}). 
 
 Another way to cross-check the expression for $G_2$ is to calculate its pole part and compare it with the heat kernel coefficient through the relation \cite{toms1}
    \begin{align}
        G_{2k}(x,x) = (4\pi)^{-N/2}(m^2)^{N/2-k-1}\Gamma(k+1-N/2)E_k(x)
    \end{align}
    where $N$ is the space-time dimension and $E_k$ are the heat kernel coefficients which have been computed for the case of spin-2 gravity in \cite{tomsapp3}. Using the following results from dimensional regularization,
    \begin{align}
        &\int \frac{d^4k}{(2\pi)^4}\frac{1}{k^4} = \frac{1}{8\pi^2\epsilon},\nonumber\\
        &\int \frac{d^4k}{(2\pi)^4}\frac{k^\alpha k^\beta}{k^4} = 0,\nonumber\\
        &\int \frac{d^4k}{(2\pi)^4}\frac{k^\alpha k^\beta}{k^6} = \frac{g^{\alpha\beta}}{32\pi^2\epsilon},\nonumber\\
         &\int \frac{d^4k}{(2\pi)^4}\frac{k^\alpha k^\beta k^\gamma k^\tau}{k^8} = \frac{g^{\alpha\beta}g^{\gamma\tau}+g^{\alpha\gamma}g^{\beta\tau}+g^{\alpha\tau}g^{\beta\gamma}}{192\pi^2\epsilon},
    \end{align}
    where $\varepsilon = 4-N$, we find the pole part of (\ref{G2g}) taking $y$ as the origin,
    \begin{align}
        \text{Tr}[(G_2)_{\mu\nu\gamma\tau}] = -\frac{R}{2\pi^2\epsilon} + \frac{5\alpha R}{24\pi^2\epsilon} - \frac{\alpha^2R}{4\pi^2\epsilon},
    \end{align}
    which implies
    \begin{align}
        \text{Tr}\ E_1 = -4R + \frac{5\alpha R}{3}-2\alpha^2 R.\label{ourtre1}
    \end{align}

    At this stage, we claim that the expression for Tr $E_1$ has been misprinted or evaluated incorrectly in \cite{tomsapp3} (Eq. 4.42 therein). The expression for Tr $E_1$ given there (Eq. 4.41) reads,
    \begin{align}
        \text{Tr}\ E_1 = &g^{\mu\nu}g_{\rho\sigma}Q_{\mu\nu}^{\ \ \rho\sigma}\frac{(1+\zeta)^{-N/2}-1}{N} + \delta_{\rho\sigma}^{\ \ \mu\nu}Q_{\mu\nu}^{\ \ \rho\sigma}\frac{2-N-2(1+\zeta)^{-N/2}}{N}\nonumber\\
        &+R\Bigg\{\frac{N^3-N^2-12N-48}{12N}+\frac{(1+\zeta)N^2+6(1+\zeta)N+24}{6N}(1+\zeta)^{-N/2}\Bigg\},\label{tre1}
    \end{align}
    where $N$ is the space-time dimension and $\zeta$ is the gauge parameter. The relation between the gauge parameter $\zeta$ and the one appearing in this work is,
    \begin{align}
        1+\zeta = \frac{1}{\alpha}.
    \end{align}
    For Einstein gravity without cosmological constant the expression for $Q_{\mu\nu}^{\ \ \lambda\tau}$ is,
    \begin{align}
        Q_{\mu\nu}^{\ \ \lambda\tau} = &R^{\lambda\ \ \tau}_{\ \mu\nu} + R^{\lambda\ \ \tau}_{\ \nu\mu}-\frac{1}{2}(\delta^{\lambda}_{\mu}R^\tau_{\ \nu}+\delta^{\lambda}_{\nu}R^\tau_{\ \mu}+\delta^{\tau}_{\mu}R^\lambda_{\ \nu}+\delta^{\tau}_{\nu}R^\lambda_{\ \mu}) + g^{\lambda\tau}R_{\mu\nu}\nonumber\\
        &+\frac{2}{m-2}g_{\mu\nu}\left(R^{\lambda\tau}-\frac{1}{2}Rg^{\lambda\tau}\right) + R\delta^{\lambda\tau}_{\mu\nu}\label{qm}.
    \end{align}
    Using this in (\ref{tre1}) we find the correct result to be 
    \begin{align}
        \text{Tr}\ E_1 = R\Bigg\{\frac{5N^3-27N^2+70N-72}{12(2-N)}+\frac{18-5N+(N+6)\zeta}{6}(1+\zeta)^{-N/2}\Bigg\},
    \end{align}
    in place of Eq (4.42)  of \cite{tomsapp3}.
    Setting $1 + \zeta = 1/\alpha$ and $N=4$ in the expression above leads to the correct result for Tr $E_1$ as found in (\ref{ourtre1}). Another reason why (\ref{ourtre1}) has to be correct is that when $\alpha\rightarrow 1$ (or equivalently $\zeta\rightarrow 0$) limit is taken, the operator turns into minimal form for which the expression for Tr $E_1$ is well known (see \cite{hamidew, gilkey, dewittdynamical}),
    \begin{align}
        \text{Tr}\ E_1 = \text{Tr}(\frac{R}{6}\mathbf{1}-\mathbf{Q}).
    \end{align}
    Using (\ref{qm}) we find
    \begin{align}
        \text{Tr}\ E_1 = -\frac{13R}{3},
    \end{align}
    which agrees with (\ref{ourtre1}) when $\alpha\rightarrow 1$ limit is taken.
    \section{One-loop divergences in the VD effective action}\label{SecIV}
  We are interested here in obtaining the divergences in the averages $\langle S_1\rangle$,  $\langle S_2\rangle$, and  $\langle S_1^2\rangle$. This is done by using the expressions for propagators in place of averages over quantum fields in momentum space and then computing the divergences in the loop integrals that follow. We employ dimensional regularisation for this task.
  We begin with $S_2$ (Eq.(\ref{s2})). This involves the following terms whose divergent parts have been written below.
\begin{align}
    &\langle\psi(x)^2\rangle= \dfrac{1}{8\pi^2\varepsilon}\left\{-m^2 + R\left(\dfrac{1}{6}-\dfrac{\omega}{2}-\xi(1-\omega)\right)\right\},\nonumber\\
    &\langle h_{\alpha\beta}(x)h_{\mu\nu}(x)\rangle = \dfrac{1}{\varepsilon}\Bigg\{\dfrac{1}{48\pi^2}\left(\alpha(6+\alpha-2\omega(1+\alpha))-2\omega-1\right)\left(\bar g_{\alpha(\mu} R_{\nu)\beta}+\bar g_{\beta(\mu} R_{\nu)\alpha}\right)\nonumber\\
    &\hspace{30mm} +\dfrac{\alpha(\omega-2)}{16\pi^2}(\bar g_{\alpha\beta}R_{\mu\nu}+\bar g_{\mu\nu}R_{\alpha\beta})+\dfrac{1}{96\pi^2}(3\alpha(2-\omega)-1)\bar g_{\alpha\beta}\bar g_{\mu\nu}R\nonumber\\&\hspace{30mm}+\dfrac{1}{48\pi^2}\left(\alpha(\omega-2\alpha-1)+\omega-2\right)\bar g_{\alpha(\mu}\bar g_{\nu)\beta}R-\dfrac{(1+\alpha)}{8\pi^2}R_{\alpha(\mu\nu)\beta}\Bigg\},\nonumber\\
    &\langle\psi(x)\nabla_\mu\nabla_\nu\psi(x)\rangle = \dfrac{m^2}{96\pi^2\varepsilon}\left\{(1-3\omega-6\xi(1-\omega))\bar g_{\mu\nu}R-2R_{\mu\nu}-3m^2\bar g_{\mu\nu}\right\},\label{s2div}
\end{align}
        
        where, as stated earlier, wherever the derivative on the quantum field appears the derivative of the propagator is evaluated first and then the limit $x\rightarrow y$ is taken. All other terms in $\langle S_2\rangle$ (\ref{s2}) vanish within the scheme of dimensional regularisation.

    The divergent part of $\langle S_1^2\rangle$ has been written down in Appendix \ref{AppD}. The important thing to note is that the expressions for divergences appear in normal coordinates after the loop integrals are evaluated. These must be written down in a generally covariant form taking care that the final expression is in a correct tensor form. These expressions can be interpreted as covariant derivatives acting on $\delta$-functions\cite{curvasymp}. Up to fourth order derivatives appear in $\langle S_1^2\rangle$ and the relevant expressions for the same have been conveniently collected in the Appendix \ref{AppE}.

    The ghost operator up to the second order in the background scalar field reads,
    \begin{align}
        Q_{\alpha\beta}=g_{\alpha\beta}\Box-R_{\alpha\beta}-\dfrac{\kappa^2 g_{\alpha\beta}}{2}\xi\bar\phi\nabla^\mu\bar\phi\nabla_\mu+\dfrac{\kappa^2\xi}{2}\nabla_\alpha(\bar\phi\nabla_\beta\bar\phi)-\dfrac{\kappa^2}{2}\nabla_\alpha\bar\phi\nabla_\beta\bar\phi.
    \end{align}
    After redefining $\nabla_\mu\rightarrow\nabla_\mu+\dfrac{1}{4}\xi\bar\phi\nabla_\mu\bar\phi$, we can rewrite the operator in a minimal form (again up to $\mathcal{O}(\bar\phi^2)$),
    \begin{align}
         Q_{\alpha\beta}=g_{\alpha\beta}\Box-R_{\alpha\beta}+\dfrac{\kappa^2g_{\alpha\beta}}{4}\xi\nabla^\mu(\bar\phi\nabla_\mu\bar\phi)+\dfrac{\kappa^2\xi}{2}\nabla_\alpha(\bar\phi\nabla_\beta\bar\phi)-\dfrac{\kappa^2}{2}\nabla_\alpha\bar\phi\nabla_\beta\bar\phi.
    \end{align}
   A straightforward use of the heat kernel method gives us the following expression for the divergence in the ghost term,
   \begin{align}
        -\text{ln Det}Q^\alpha{}_\beta = \dfrac{\kappa^2}{16\pi^2\varepsilon}\left(\dfrac{1}{6}R\bar\phi\Box\bar\phi-R^{\alpha\beta}\nabla_\alpha\bar\phi\nabla_\beta\bar\phi\right)
   \end{align}
   After some simplifications, we end up with the following expression for divergences in one-loop effective action \ref{finaleffac},
   \begin{align}                            \Gamma^{(1)}|_{\text{div}}=\int d^4x\sqrt{g}\Big(A\bar\phi\Box^2\bar\phi+B\bar\phi\Box\bar\phi+C\bar\phi^2+D\bar\phi^2R+ER\bar\phi\Box\bar\phi+FR^{\mu\nu}\bar\phi\nabla_\mu\nabla_\nu\bar\phi\Big),\label{diveffac}
   \end{align}
   where,
   \begin{align}\label{gencoeff}
   &A = \dfrac{\kappa^2}{16\pi^2\varepsilon}\left(-\dfrac{1}{2}-\omega+\dfrac{\alpha\omega}{2}+\dfrac{3\omega^2}{8}-\dfrac{\alpha\omega^2}{8}\right),\nonumber\\
   &B = \dfrac{m^2\kappa^2}{16\pi^2\varepsilon}\left(-1+\dfrac{\alpha}{2}+\dfrac{3\xi^2}{2}+\dfrac{19\omega}{8}-\alpha\omega-\dfrac{\xi\omega}{4}-\dfrac{3\xi^2\omega}{2}-\dfrac{3\omega^2}{4}+\dfrac{\alpha\omega^2}{4}\right),\nonumber\\
   &C=\dfrac{m^4\kappa^2}{16\pi^2\varepsilon}\left(\dfrac{3}{2}-\dfrac{\alpha}{2}-2\xi+\dfrac{3\xi^2}{2}-\dfrac{5\omega}{4}+\dfrac{\alpha\omega}{2}+\dfrac{3\xi^2\omega}{2}+\dfrac{3\omega^2}{8}-\dfrac{\alpha\omega^2}{8}\right)-\dfrac{m^2\lambda}{32\pi^2\varepsilon},\nonumber\\
   &D=\dfrac{m^2\kappa^2}{16\pi^2\varepsilon}\Bigg(1-\dfrac{5\alpha}{12}+\dfrac{\alpha^2}{2}+\dfrac{11\xi}{6}-\dfrac{\alpha\xi}{2}-\dfrac{13\xi^2}{4}+3\xi^3-\dfrac{\omega}{24}-\dfrac{13\xi\omega}{4}+\dfrac{11\alpha\xi\omega}{12}-\dfrac{\alpha^2\xi\omega}{2}\nonumber\\&\hspace{16mm}+\frac{7\xi^2\omega}{2}+\dfrac{\omega^2(1+\xi-\alpha\xi)}{8}+\dfrac{3\xi^2\omega^2}{4}-\dfrac{3\xi^3\omega}{2}-\dfrac{3\xi^3\omega^2}{2}\Bigg)+\dfrac{\lambda}{32\pi^2\varepsilon}\left(\dfrac{1}{6}-\xi-\dfrac{\omega}{2}+\xi\omega\right),\nonumber\\
   &E=\dfrac{\kappa^2}{16\pi^2\varepsilon}\Bigg(-\dfrac{7}{12}+\dfrac{5\alpha}{6}-\dfrac{\alpha^2}{6}-\dfrac{\xi}{2}-\dfrac{\xi^2}{4}+\dfrac{3\xi^3}{2}+\dfrac{31\omega}{48}-\dfrac{19\alpha\omega}{24}-\dfrac{\alpha^2\omega}{12}+\dfrac{17\xi\omega}{12}-\dfrac{5\alpha\xi\omega}{12}\nonumber\\&\hspace{16mm}+\dfrac{\alpha^2\xi\omega}{2}+\dfrac{3\xi^2\omega}{4}-3\xi^3\omega+\dfrac{\omega^2}{48}+\dfrac{\alpha\omega^2(1+\alpha)}{12}-\dfrac{3\xi\omega^2}{8}+\dfrac{\alpha\xi\omega^2}{8}-\dfrac{\xi^2\omega^2}{2}+\dfrac{3\xi^3\omega^2}{2}\Bigg),\nonumber\\
   &F=\dfrac{\kappa^2}{16\pi^2\varepsilon}\Bigg(-\dfrac{7}{6}-\dfrac{\alpha}{2}-\dfrac{\alpha^2}{3}+2\xi-\dfrac{\omega}{2}+\dfrac{13\alpha\omega}{16}+\dfrac{\alpha^2\omega}{3}-\xi\omega-\dfrac{\omega^2(1+\alpha+\alpha^2)}{3}\Bigg),
   \end{align}
   where $\varepsilon = 4-N$. The expression for gauge-independent effective action can be derived by taking the limit $\alpha\rightarrow 0$ and $\omega\rightarrow 1$. The coefficients now read,
   \begin{align}
   &A = -\dfrac{9\kappa^2}{128\pi^2\varepsilon},\nonumber\\
   &B = \dfrac{m^2\kappa^2}{64\pi^2\varepsilon}\left(\dfrac{5}{2}-\xi\right),\nonumber\\
   &C=\dfrac{m^4\kappa^2}{16\pi^2\varepsilon}\left(\dfrac{5}{8}-2\xi+3\xi^2\right)-\dfrac{m^2\lambda}{32\pi^2\varepsilon},\nonumber\\
   &D=\dfrac{m^2\kappa^2}{32\pi^2\varepsilon}\left(\dfrac{13}{6}-\dfrac{31\xi}{12}+2\xi^2\right)-\dfrac{\lambda}{96\pi^2\varepsilon},\nonumber\\
   &E=\dfrac{\kappa^2}{192\pi^2\varepsilon}\left(1+\dfrac{13\xi}{2}\right),\nonumber\\
   &F=\dfrac{\kappa^2}{16\pi^2\varepsilon}(\xi-2).\label{vdcoeff}
   \end{align}
   A few comparisons with previous work are in order. The flat spacetime background limit ($\bar g_{\mu\nu}=\eta_{\mu\nu}$) of the expression (\ref{diveffac}) is in agreement with \cite{ourpaper1,mackaytoms}. 
   If we take the limit $\alpha\rightarrow 1$ and $\omega\rightarrow 0$, we get the standard effective action in the Landau-DeWitt gauge with the following coefficients,
   \begin{align}
        &A=-\dfrac{\kappa^2}{32\pi^2\varepsilon},\nonumber\\
        &B=\dfrac{m^2\kappa^2}{32\pi^2\varepsilon}(3\xi^2-1),\nonumber\\
        &C=\dfrac{m^4\kappa^2}{16\pi^2\varepsilon}\left(1-2\xi+\dfrac{3\xi^2}{2}\right)-\dfrac{m^2\lambda}{32\pi^2\varepsilon},\nonumber\\
        &D=\dfrac{m^2\kappa^2}{16\pi^2\varepsilon}\left(\dfrac{13}{12}+\dfrac{4\xi}{3}-\dfrac{13\xi^2}{4}+3\xi^3\right)+\dfrac{\lambda}{32\pi^2\varepsilon}\left(\dfrac{1}{6}-\xi\right),\nonumber\\
        &E=\dfrac{\kappa^2}{32\pi^2\varepsilon}\left(\dfrac{1}{6}-\xi-\dfrac{\xi^2}{2}+3\xi^3\right),\nonumber\\
        &F=\dfrac{\kappa^2}{8\pi^2\varepsilon}(\xi-1).\label{scoeff}
   \end{align}
   On comparing the expression for effective action (\ref{diveffac}) with coefficients (\ref{scoeff}) against \cite{standard}, where the authors have carried
out similar computations for a general Lagrangian for multiple scalar fields coupled with gravity in the Jordan frame we find that the result here is in disagreement because the authors have used a different gauge condition showing that the standard effective action is indeed gauge dependent when computed off-shell for gauge theories. However, if we take a constant scalar field background $\nabla_\mu\bar\phi = 0$ then the expression agrees with \cite{standard} since then the gauge conditions become identical.
\section{Renormalization and comparison of the effective actions}\label{SecV}
The expression for gauge-independent divergences in one-loop effective action is given by (\ref{diveffac}) with coefficients given in (\ref{vdcoeff}). The divergences may be removed with the following counter-terms,
\begin{align}
    &\delta Z_\phi = \dfrac{m^2\kappa^2}{32\pi^2\varepsilon}\left(\dfrac{5}{2}-\xi\right),\nonumber\\
    &\delta m^2 = \dfrac{m^4\kappa^2}{8\pi^2\varepsilon}\left(\dfrac{9\xi}{4}-\dfrac{5}{4}-3\xi^2\right)+\dfrac{m^2\lambda}{16\pi^2\varepsilon},\nonumber\\
    &\delta\xi = \dfrac{m^2\kappa^2}{32\pi^2\varepsilon}\left(\dfrac{8\xi}{3}-\dfrac{13}{3}-3\xi^2\right)+\dfrac{\lambda}{48\pi^2\varepsilon}.\label{vdct}
\end{align}
We also need to add new terms in the classical action to remove the divergences $ER\bar\phi\Box\bar\phi$ and $FR^{\mu\nu}\bar\phi\nabla_\mu\nabla_\nu\bar\phi$ in the effective action.

Renormalization of coupling constants from the standard effective action in Landau-DeWitt gauge (\ref{diveffac}) with coefficients (\ref{scoeff}) read,
\begin{align}
    &\delta Z_\phi = \dfrac{m^2\kappa^2}{16\pi^2\varepsilon}(3\xi^2-1),\nonumber\\
    &\delta m^2 = \dfrac{m^4\kappa^2}{4\pi^2\varepsilon}\left(\xi-\dfrac{1}{4}-\dfrac{3\xi^2}{2}\right) + \dfrac{m^2\lambda}{16\pi^2\varepsilon},\nonumber\\
    &\delta\xi = \dfrac{m^2\kappa^2}{16\pi^2\varepsilon}\left(\dfrac{13\xi^2}{2}-\dfrac{5\xi}{3}-\dfrac{13}{6}-9\xi^3\right)+\dfrac{\lambda}{16\pi^2\varepsilon}\left(\xi-\dfrac{1}{6}\right).\label{sct}
\end{align}
Let us check whether the effective actions match in the limit $\kappa=0$. We find in the limit $\kappa=0$,
\begin{equation}\label{vdctk0}
  \left.\begin{aligned}
  &\delta m^2 = \dfrac{m^2\lambda}{16\pi^2\varepsilon}\\
  &\delta \xi = \dfrac{\lambda}{48\pi^2\varepsilon}\\
\end{aligned}\right\}  \text{VD}
\end{equation}
\begin{equation}\label{sctk0}
  \left.\begin{aligned}
  &\delta m^2 = \dfrac{m^2\lambda}{16\pi^2\varepsilon}\\
  &\delta \xi = \dfrac{\lambda}{16\pi^2\varepsilon}\left(\xi-\dfrac{1}{6}\right)\\
\end{aligned}\right\}  \text{Standard}
\end{equation}
We can see that the difference lies in the renormalization of $\xi$. To make any further statement in this regard we must look closely at the expression for coefficient $D$ in Eqs.~(\ref{gencoeff}). We can see that the difference in the expressions for $D$ in Eqs.~(\ref{vdcoeff}) and Eqs.~(\ref{scoeff}) in $\kappa = 0$ limit arises after setting the value for $\omega$ in the expression for $D$ in Eqs.~(\ref{gencoeff}) as shown below.
\begin{align}
    &\frac{\lambda}{32\pi^2\varepsilon}\left(\frac{1}{6}-\xi-\frac{\omega}{2}+\xi\omega\right)\xrightarrow[]{\omega=1}-\frac{\lambda}{96\pi^2\varepsilon} \hspace{8mm}[\text{VD}],\\
    &\frac{\lambda}{32\pi^2\varepsilon}\left(\frac{1}{6}-\xi-\frac{\omega}{2}+\xi\omega\right)\xrightarrow[]{\omega=0}\frac{\lambda}{32\pi^2\varepsilon}\left(\frac{1}{6}-\xi\right)\hspace{8mm}[\text{standard}].
\end{align}
This makes it clear that the difference in renormalization for $\xi$ arises due to the field space Christoffel connections. We demonstrate below that this is indeed the case.

In what follows we will focus entirely on the terms that involve $R\bar\phi^2$ in $\kappa\rightarrow 0$ limit. Such terms arise from $N_1(\bar\phi)\psi^2$ in the expression for $S_2$ (\ref{s2}). The terms not involving $\omega$ in $N_1(\bar\phi)\psi^2$ read (see Appendix \ref{AppB})
\begin{align}
    \left\{\frac{\lambda}{4}\bar\phi^2+\dfrac{1}{2}\omega\bar\phi^2\left(m^2\xi-\dfrac{m^2}{4}-\dfrac{3m^2\xi^2}{2}+\frac{\xi^2R}{4}\right)\right\}\langle\psi^2\rangle \xrightarrow[]{\text{not involving $\omega$}} \frac{\lambda}{32\pi^2\varepsilon}R\bar\phi^2\left(\frac{1}{6}-\xi\right)
\end{align}
which agrees with the standard result. The terms involving $\omega$ in $N_1(\bar\phi)\psi^2$ read,
\begin{align}
    \left\{\frac{\lambda}{4}\bar\phi^2+\dfrac{1}{2}\omega\bar\phi^2\left(m^2\xi-\dfrac{m^2}{4}-\dfrac{3m^2\xi^2}{2}+\frac{\xi^2R}{4}\right)\right\}\langle\psi^2\rangle \xrightarrow[]{\text{involving $\omega$}} \frac{\lambda\omega}{32\pi^2\varepsilon}R\bar\phi^2\left(\xi-\frac{1}{2}\right)
\end{align}
which is completely independent of $\kappa$ causing the discrepancy between the VD and the standard effective actions.

But how do the field space connections which are all proportional to $\kappa$ (see Appendix \ref{AppA}), contribute towards the effective action through terms which are independent of $\kappa$? To see this we consider the following term involving the field space connections.
\begin{align}
    -\frac{\omega}{2}(\Gamma^1_{22})_{\mu\nu}(S_{,1})^{\mu\nu}\psi^2
\end{align}
We remind the readers of the convention we introduced following Eq.(\ref{gen}) that is being used here. The term mentioned above originates from $\Gamma^k_{ij}S_{,k}\eta^i\eta^j$ in Eq.~\ref{Sq}. The expression for $(\Gamma^1_{22})_{\mu\nu}$ from the Appendix \ref{AppA} reads.
\begin{align}\label{gamma122}
    (\Gamma^1_{22})_{\mu\nu}(x,x',x'') =\dfrac{\kappa }{4}\tilde{\delta}(x'',x')\tilde{\delta}(x'',x)\Bigg\{1-2\xi+\frac{\kappa^2\xi\bar\phi^2(x)}{4}\left(\frac{1}{4}-\xi+3\xi^2\right)\Bigg\}\bar g_{\mu\nu}(x)
\end{align}
And the derivative of the action wrt. the metric reads,
\begin{align}\label{dLg}
    (S_{,1})^{\mu\nu}(x) &= \sqrt{|g|}\kappa\Bigg(\frac{1}{4}m^2\bar g^{\mu\nu}(x)\bar\phi^2(x)-\frac{\xi}{2}\left(R^{\mu\nu}(x)-\frac{1}{2}\bar g^{\mu\nu}(x)R(x)\right)\bar\phi^2(x)+\left(\xi-\frac{1}{2}\right)\nabla^\mu\bar\phi(x)\nabla^\nu\bar\phi(x)\nonumber\\&\hspace{15mm}+\xi\bar\phi(x)\nabla^\mu\nabla^\nu\bar\phi(x)+\left(\frac{1}{4}-\xi\right)\bar g^{\mu\nu}(x)\nabla_\alpha\bar\phi(x)\nabla^\alpha\bar\phi(x)-\xi \bar g^{\mu\nu}(x)\bar\phi(x)\Box\bar\phi(x)\Bigg)\nonumber\\&\hspace{15mm}+\frac{\sqrt{|g|}}{\kappa}(2R^{\mu\nu}(x)-\bar g^{\mu\nu}(x)R(x)).
\end{align}
The product of Eqs.~(\ref{gamma122}) and (\ref{dLg}) would then contain terms independent of $\kappa$,
\begin{align}
    -\frac{\omega}{2}(\Gamma^1_{22})_{\mu\nu}(S_{,1})^{\mu\nu}\psi^2 = \left(\frac{\omega}{2}-\omega\xi\right)R\psi^2 + \text{other $\kappa$-dependent terms} 
\end{align}
These are precisely the terms appearing in the equation of the scalar field propagator (\ref{seq}). Since these connection-dependent terms in the equation of the scalar field propagator also depend on the background curvature, the discrepancy between the effective actions in the limit $\kappa=0$ is absent when they are computed for a flat Minkowskian or Euclidean metric background. (This is why the effective actions match in the limit $\kappa =0$ in \cite{mackaytoms}) 

Now that we have shown that the effective actions differ in the limit $\kappa=0$, let us also see how the effective actions compare if we start with the classical theory \ref{action} without the Ricci scalar term. The classical theory is now that of a scalar field non-minimally coupled to the background curvature. In that case, the field space metric has a single component that reads,
\begin{align}
    G_{22} = \sqrt{|\bar g|}\tilde \delta(x,x')
\end{align}
Since the field space metric is independent of the fields, the Christoffel connections associated with the field space metric vanish. The VD effective action, in this case, corresponds to Eq.~\ref{diveffac}
with $\omega$ taken zero in \ref{gencoeff} (of course, we throw away all the $\kappa$-dependent terms in the absence of gravity ($h_{\mu\nu}=0$)). Since the $\kappa$-independent terms in \ref{gencoeff} are also independent of the gauge parameter $\alpha$ it is clear that the standard effective action \cite{tomsbook, standard} matches the VD effective action as is expected for a non-gauge theory. 

Let us summarise what we have found. We have found that the effective actions do not match in the limit $\kappa=0$, taken after computing them for the classical theory \ref{action} contrary to the result of \cite{mackaytoms}. However, the effective actions for the same theory match if the spacetime metric is only a background field. Thus, taking the limit $\kappa=0$ after computing the VD effective is NOT equivalent to computing the VD effective action for the same theory where the spacetime metric is a background field.

\section{Conclusion}
We performed the calculation of one-loop divergences in the Vilkovisky-DeWitt(VD) effective action for the theory of a scalar field non-minimally coupled to gravity. We assumed an arbitrary background for both the scalar field and the metric. The curvature expansion of the propagators is obtained through the use of the technique of local momentum space. The expression for the scalar field propagator is shown to match the results obtained earlier when the connection-dependent terms are switched off. In particular, the expression for graviton propagator is used to obtain the heat kernel coefficient $E_1$. In doing so we have corrected the result of \cite{tomsapp3}.

Because the VD effective action is gauge-independent off-shell and therefore different from the standard effective action for gauge theories, it is expected to match the standard effective action even when computed off-shell for a non-gauge theory like that of a scalar field where the spacetime metric is only a background field. However, in the presence of gravity, when the VD effective action is computed it does not match the standard effective in the limit $\kappa=0$. We have shown that a non-trivial cancellation of $\kappa$ factors among the connection and the functional derivative of the action was found to be the cause of the discrepancy. From this, we conclude that computing the VD effective action for the theory \ref{action} and taking the limit $\kappa=0$ is NOT the same as starting with the same theory in the absence of gravity. This is unlike what was found in \cite{mackaytoms}, since the discrepancy is caused by terms that depend on the background curvature. As a result, there is no discrepancy between the standard and the VD effective action if the background metric is chosen to be either a flat Minkowskian or Euclidean.
  \section*{Acknowledgement}
The calculations in this paper have been carried out in MATHEMATICA using the xAct packages xTensor \cite{xtensor} and xPert \cite{xpert}. This work is partially supported by the DST (Govt. of India) Grant
    No. SERB/PHY/2021057.
\appendix
\section{Field space Christoffel connections}\label{AppA}
The components of the inverse field-space metric are given by the following expressions.

\begin{align}
    &G^{11}_{\mu\nu\alpha\beta}(x,x') = \Bigg\{\frac{1}{2}\left(1+\frac{\xi\bar\phi^2(x)}{4}\right)(2\bar g_{\mu(\alpha}(x)\bar g_{\beta)\nu}(x)-\bar g_{\mu\nu}(x)\bar g_{\alpha\beta}(x))+\frac{\xi^2\bar\phi^2(x)}{4}\bar g_{\mu\nu}(x)\bar g_{\alpha\beta}(x)\Bigg\}\delta(x,x'),\nonumber\\
    &G^{12}_{\mu\nu}(x,x')=\frac{\xi\bar\phi(x)}{2}\bar g_{\mu\nu}(x)\delta(x,x'),\nonumber\\
    &G^{22}(x,x')=\Big(1-\frac{3\xi^2\bar\phi^2(x)}{2}\Big)\delta(x,x'),
\end{align}

The Christoffel connection $\Gamma^i_{jk}$ arising from the field space metric is given by the following expression,
\begin{align}
    \Gamma^i_{jk} = \frac{1}{2}g^{il}(g_{lj,k}+g_{lk,j}-g_{jk,l})
\end{align}
A straightforward calculation of the components of the connections yields,
\begin{align}
   &( \Gamma^1_{11})^{ \mu\nu\rho\sigma}_{\lambda\tau}(x,x',x'') = \kappa\tilde{\delta}(x'',x')\tilde{\delta}(x'',x)\Bigg(-\delta^{(\mu}_{(\lambda}\bar g^{\nu)(\rho}(x)\delta^{\sigma)}_{\tau)} +\dfrac{1}{4}\bar g^{\mu\nu}(x)\delta^{\rho}_{(\lambda}\delta^{\sigma}_{\tau)}+\dfrac{1}{4}\bar g^{\rho\sigma}(x)\delta^{\mu}_{(\lambda}\delta^{\nu}_{\tau)} \nonumber\\&\hspace{35 mm}+\dfrac{1}{4}\left(1-\frac{\kappa^2\xi^2\bar\phi^2(x)}{2}\right)\bar g_{\lambda\tau}(x)\bar g^{\mu(\rho}(x)\bar g^{\sigma)\nu}(x) -\dfrac{1}{8}\left(1-\frac{\kappa^2\xi^2\bar\phi^2(x)}{2}\right)\bar g_{\lambda\tau}(x)\bar g^{\mu\nu}(x)\bar g^{\rho\sigma}(x)\Bigg) \nonumber\\
  &(\Gamma^1_{12})^{\mu\nu}_{\rho\sigma}(x,x',x'') = \dfrac{\kappa^2\xi\bar\phi(x)}{8}\tilde{\delta}(x'',x')\tilde{\delta}(x'',x)(\bar g_{\rho\sigma}(x)\bar g^{\mu\nu}(x) - 2\delta^\mu_{(\rho}\delta^\nu_{\sigma)})\nonumber\\
  &(\Gamma^1_{22})_{\mu\nu}(x,x',x'') =\dfrac{\kappa }{4}\tilde{\delta}(x'',x')\tilde{\delta}(x'',x)\Bigg(1-2\xi+\frac{\kappa^2\xi\bar\phi^2(x)}{4}\left(\frac{1}{4}-\xi+3\xi^2\right)\Bigg)\bar g_{\mu\nu}(x) \nonumber\\
  &(\Gamma^2_{11})^{\mu\nu\rho\sigma}(x,x',x'')=\dfrac{\kappa^2\xi\bar\phi(x)}{8}\tilde{\delta}(x'',x')\tilde{\delta}(x'',x)(\bar g^{\mu\nu}(x)\bar g^{\rho\sigma}(x)-2\bar g^{\mu(\rho}(x)\bar g^{\sigma)\nu}(x))\nonumber\\
  &(\Gamma^2_{12})^{\mu\nu}(x,x',x'')=\dfrac{\kappa}{4}\tilde{\delta}(x'',x')\tilde{\delta}(x'',x) \bar g^{\mu\nu}(x) \nonumber\\
  &(\Gamma^2_{22})(x,x',x'')=\dfrac{\kappa^2\xi\bar\phi(x)}{2}\tilde{\delta}(x'',x')\tilde{\delta}(x'',x)\left(3\xi-1\right)
\end{align}
\section{Functions appearing in the expression for $S_1$ and $S_2$}\label{AppB}
The expression for $S_1$ contains terms that are of the first order in the background scalar field. The various functions that appear in the expression for $S_1$ \ref{s1} read,
\begin{align}
    &M_1^{\mu\nu}(\bar\phi)=\dfrac{1}{2}\Big\{\bar\phi \bar g^{\mu\nu}\left(1-\dfrac{\omega}{2}\right)(m^2+\xi R)-2\xi\bar\phi R_{\alpha\beta}\bar g^{\alpha(\mu}\bar g^{\nu)\beta}\left(1-\dfrac{\omega}{2}\right)\nonumber\\&\hspace{10mm}+\dfrac{\xi-1}{\alpha}(\bar g^{\mu\nu}\Box\bar\phi-2\bar g^{\alpha(\mu}\nabla_\beta\nabla^{\nu)}\bar\phi)+2\xi \bar g^{\alpha(\mu}\bar g^{\nu)\beta}\nabla_\alpha\nabla_\beta\bar\phi+\bar g^{\mu\nu}\Box\bar\phi\left(\dfrac{\omega}{2}-2\xi\right)\Big\}\nonumber \\
    &M_2^{\mu\nu\alpha}(\bar\phi)=\bar g^{\alpha(\mu}\nabla^{\nu)}\bar\phi\left(\dfrac{1}{\alpha}(1-\xi)-1\right)+\bar g^{\mu\nu}\nabla^\alpha\bar\phi\left(\dfrac{1}{2\alpha}(\xi-1)-1\right)\nonumber\\
    &\hspace{20mm}+\xi \bar g^{\alpha(\mu}\bar g^{\nu)\beta}\bar\phi\left(2-\dfrac{1}{\alpha}\right)+\bar g^{\mu\nu}\bar g^{\alpha\beta}\nabla_\beta\bar\phi\left(\xi\left(\dfrac{1}{2\alpha}-1\right)+\dfrac{1}{2}\right)\\
    &M_3^{\mu\nu\alpha\beta}(\bar\phi) = \xi \bar g^{\mu(\alpha}\bar g^{\beta)\nu}\bar\phi\left(1-\dfrac{1}{\alpha}\right)+\xi \bar g^{\alpha\beta}\bar g^{\mu\nu}\bar\phi\left(\dfrac{1}{\alpha}-2\right)
\end{align}
The expression for $S_2$ contains terms that are quadratic in the background scalar field. The various functions that appear in the expression for $S_2$ \ref{s1} read,
\begin{align}
    &N_1(\bar\phi)=\frac{\lambda}{4}\bar\phi^2+\dfrac{1}{2}\omega\bar\phi^2\left(m^2\xi-\dfrac{m^2}{4}-\dfrac{3m^2\xi^2}{2}+\frac{\xi^2R}{4}\right)+\dfrac{\xi\bar\phi\Box\bar\phi}{2}\left(\dfrac{1}{\alpha}-\dfrac{\xi}{\alpha}+\dfrac{\omega}{4}\right)\nonumber\\&\hspace{15mm}+\dfrac{\nabla_\alpha\bar\phi\nabla^\alpha\bar\phi}{2}\left(\dfrac{1}{2\alpha}-\dfrac{\xi}{2\alpha}-\dfrac{\omega}{8}+\xi\omega-\dfrac{3\xi^2\omega}{2}\right)\\
    &N_2^\mu(\bar\phi)=\dfrac{\xi\bar\phi\nabla^\nu\bar\phi}{\alpha}\left(\dfrac{1}{2}-\xi\right)\\
    &N_3^{\mu\nu}(\bar\phi)=-\dfrac{\xi^2\bar g^{\mu\nu}\bar\phi^2}{4\alpha}\\
    &N_4^{\mu\nu\alpha\beta}(\bar\phi)=-\dfrac{\xi}{8}\left(1-\dfrac{\omega}{2}\right)\bar\phi^2(\bar g^{\mu\nu}R^{\alpha\beta}+\bar g^{\alpha\beta}R^{\mu\nu}-2\bar g^{\alpha(\mu}R^{\nu)\beta}-2\bar g^{\beta(\mu}R^{\nu)\alpha})\nonumber\\&\hspace{10mm}+\dfrac{m^2\bar\phi^2(1-\xi\omega)}{16}(\bar g^{\alpha\beta}\bar g^{\mu\nu}-2\bar g^{\alpha(\mu}\bar g^{\nu)\beta})+\dfrac{\xi}{16}\bar\phi^2R\left(1-\dfrac{\omega}{2}-\xi\omega\right)(\bar g^{\alpha\beta}\bar g^{\mu\nu}-2\bar g^{\alpha(\mu}\bar g^{\nu)\beta})\nonumber\\&\hspace{10mm}-\dfrac{1}{8}\left(1+\dfrac{2\xi}{\alpha}-\dfrac{\omega}{2}+\xi\omega\right)(\bar g^{\alpha\beta}\nabla^\mu\bar\phi\nabla^\nu\bar\phi+\bar g^{\mu\nu}\nabla^\alpha\bar\phi\nabla^\beta\bar\phi-2\bar g^{\alpha(\mu}\nabla^\beta\nabla^{\nu)}\bar\phi-2\bar g^{\beta(\mu}\nabla^\alpha\nabla^{\mu)\bar\phi})\nonumber\\&\hspace{10mm}-\dfrac{\xi}{4}\left(\dfrac{1}{\alpha}+\dfrac{\omega}{2}\right)(\bar g^{\alpha\beta}\bar\phi\nabla^\mu\nabla^\nu\bar\phi+\bar g^{\mu\nu}\nabla^\alpha\nabla^\beta\bar\phi-2\bar g^{\alpha(\mu}\bar\phi\nabla^\beta\nabla^{\nu)}\bar\phi-2\bar g^{\beta(\mu}\bar\phi\nabla^\alpha\nabla^{\nu)}\bar\phi)\nonumber\\&\hspace{10mm}+\dfrac{\xi\omega}{8}\bar\phi\Box\bar\phi(\bar g^{\alpha\beta}\bar g^{\mu\nu}-2\bar g^{\alpha(\mu}\bar g^{\nu)\beta})+\dfrac{\nabla_\tau\bar\phi\nabla^\tau\bar\phi}{16}\left(1-\dfrac{\omega}{2}+\xi\omega\right)(\bar g^{\alpha\beta}\bar g^{\mu\nu}-2\bar g^{\alpha(\mu}\bar g^{\nu)\beta})\\
    &N_5^{\mu\nu\alpha\beta\gamma}(\bar\phi)=\dfrac{\xi}{2}\left(1+\dfrac{1}{2\alpha}\right)(\bar g^{\mu(\alpha}\bar g^{\beta)\gamma}\bar\phi\nabla^\nu\bar\phi+\bar g^{\nu(\alpha}\bar g^{\beta)\gamma}\bar\phi\nabla^\mu\bar\phi+\bar g^{\alpha(\mu}\bar g^{\nu)\gamma}\bar\phi\nabla^\beta\bar\phi+\bar g^{\nu(\alpha}\bar g^{\beta)\gamma}\bar\phi\nabla^\mu\bar\phi)\nonumber\\&\hspace{10mm}-\xi \bar g^{\mu\nu}\bar g^{\gamma(\alpha}\bar\phi\nabla^{\beta)}\bar\phi\left(1+\dfrac{1}{\alpha}\right)-\dfrac{3\xi}{4}\bar g^{\alpha(\mu}\bar g^{\nu)\beta}\bar\phi\nabla^\gamma\bar\phi+\dfrac{\xi}{4}\left(1+\dfrac{1}{2\alpha}\right)\bar g^{\alpha\beta}\bar g^{\mu\nu}\bar\phi\nabla^\gamma\bar\phi\\
    &N_6^{\mu\nu\alpha\beta\gamma\tau}(\bar\phi)=\dfrac{\xi\bar\phi^2}{8}\Bigg(\dfrac{1}{\alpha}(\bar g^{\alpha\gamma}\bar g^{\tau(\mu}\bar g^{\nu)\beta}+\bar g^{\beta\gamma}\bar g^{\tau(\mu}\bar g^{\nu)\alpha})+4\bar g^{\mu(\gamma}\bar g^{\tau)\nu}\bar g^{\alpha\beta}-\bar g^{\mu\gamma}\bar g^{\nu(\alpha}\bar g^{\beta)\tau}\nonumber\\ &\hspace{15mm}-\bar g^{\nu\gamma}\bar g^{\mu(\alpha}\bar g^{\beta)\tau}+\bar g^{\mu(\alpha}\bar g^{\beta)\nu}\bar g^{\gamma\tau}-\left(1-\dfrac{1}{2\alpha}\right)\bar g^{\mu\nu}\bar g^{\alpha\beta}\bar g^{\gamma\tau}-2\left(1+\dfrac{1}{\alpha}\right)\bar g^{\mu\nu}\bar g^{\alpha(\gamma}\bar g^{\tau)\beta}\Bigg)
\end{align}
\section{Graviton propagator}\label{AppC}
The expression for the graviton propagator in the momentum space defined in Eq.~\ref{ftp} up to the first order in the background curvature reads,
 \begin{align}
        (\tilde G_2)_{\mu\nu\gamma\tau} &= \frac{2(1-\alpha)(4-\omega+\alpha(\omega-2))}{k^8}k^\alpha k^\beta R_{\alpha\beta}\ k_{(\mu}\bar g_{\nu)(\gamma}k_{\tau)}+\frac{8(1-\alpha)}{k^8}k_{(\gamma}R_{\tau)\alpha\beta(\mu}\ k_{\nu)}k^\alpha k^\beta\nonumber\\
        &+\frac{1}{3k^6}(\bar g_{\mu\nu}\bar g_{\gamma\tau}-2\bar g_{\mu(\gamma}\bar g_{\tau)\nu})k^\alpha k^\beta R_{\alpha\beta} + \frac{(1-\alpha)(2-\omega)}{k^6}\bigg(\bar g_{\mu\nu}k^\alpha k_{(\gamma}R_{\tau)\alpha}\nonumber\\
        &-2R_{\alpha(\gamma}\bar g_{\tau)(\mu}k_{\nu)}k^\alpha +k_{\mu}k_{\nu}R_{\gamma\tau}-k_{(\gamma}\bar g_{\tau)(\mu}R_{\nu)\alpha}k^\alpha+ \bar g_{\gamma\tau}k^\alpha k_{(\mu}R_{\nu)\alpha}+ k_\gamma k_\tau R_{\mu\nu} \nonumber\\&-\bar g_{\mu\nu}k_{\gamma}k_{\tau}R-\bar g_{\gamma\tau}k_\mu k_\nu R\bigg)+ \frac{2}{k^6}\left(\frac{4\alpha}{3}-\omega\alpha^2 +\omega- \frac{4}{3}\right)k_{(\mu}R_{\nu)(\gamma}k_{\tau)}\nonumber\\
         &+\frac{1}{k^6}\left(\frac{4\alpha}{3}+\omega\alpha^2-2\alpha^2+\frac{2}{3}-\omega\right)k_{(\mu}\bar g_{\nu)(\gamma}k_{\tau)}R-\frac{2}{3k^6}(\bar g_{\mu\nu}k^\alpha k^\beta R_{\gamma\alpha\tau\beta} \nonumber\\&+2R_{\tau\alpha\beta(\mu}\bar g_{\nu)\gamma}k^\alpha k^\beta+2R_{\gamma\alpha\beta(\mu}\bar g_{\nu)\tau}k^\alpha k^\beta+\bar g_{\gamma\tau}k^\alpha k^\beta R_{\mu\alpha\nu\beta}) \nonumber\\
         & + \frac{16(1-\alpha)}{3k^6}(k^\alpha R_{\alpha(\gamma\tau)(\mu} k_{\nu)}+k^\alpha k_{(\gamma}R_{\tau)(\mu\nu)\alpha})\nonumber\\
         &+\frac{1}{k^4}\left(\frac{\omega}{2}-\frac{5}{6}\right)(\bar g_{\mu\nu}R_{\gamma\tau}+\bar g_{\gamma\tau}R_{\mu\nu})+\frac{1}{3k^4}\left(\frac{2}{3}-\omega\right)(\bar g_{\mu(\gamma}R_{\tau)\nu}+\bar g_{\nu(\gamma}R_{\tau)\mu})\nonumber\\&+\frac{1}{4k^4}\left(\frac{4}{3}-\omega\right)(\bar g_{\mu\nu}\bar g_{\gamma\tau}-2\bar g_{\mu(\gamma}\bar g_{\tau)\nu})R+ \frac{2(2\alpha-5)}{3k^4}R_{\gamma(\mu\nu)\tau}\nonumber\\
         & +\dfrac{i(1-\alpha)}{3k^4}(k_{(\mu}R_{\nu)\gamma\beta\tau}+k_{(\mu}R_{\beta\gamma\nu)\tau}+4k_{(\gamma}R_{\tau)(\mu\nu)\beta}-2k_{(\mu}\bar g_{\nu)(\gamma}R_{\tau)\beta}\nonumber\\&-k^\alpha k_{(\gamma}R_{\tau)(\alpha\mu)\beta}k_\nu -k^\alpha k_{(\gamma}R_{\tau)(\alpha\mu)\beta}k_\mu)y^\beta+\dfrac{2i}{3k^4}\bar g_{\gamma\tau}k^\alpha R_{\alpha(\mu \nu)\beta}y^\beta\nonumber\\&+\dfrac{4i}{3k^4}k^\alpha(\bar g_{\mu(\gamma}R_{\tau)(\alpha\nu)\beta}+\bar g_{\nu(\gamma}R_{\tau)(\alpha\mu)\beta})y^\beta+\dfrac{(\alpha-1)}{3k^4}k_{(\mu}\bar g_{\nu)(\gamma}k_{\tau)}R_{\alpha\beta}y^\alpha y^\beta\nonumber\\&+\dfrac{2(\alpha-1)}{3k^4}k_{(\gamma}R_{\tau)\alpha\beta(\mu}k_{\nu)}y^\alpha y^\beta+\dfrac{1}{12k^2}R_{\alpha\beta}y^\alpha y^\beta(2\bar g_{\gamma(\mu}\bar g_{\nu)\tau}-\bar g_{\gamma\tau}\bar g_{\mu\nu})\nonumber\\&+\dfrac{1}{6k^2}(\bar g_{\gamma\tau}R_{\mu\alpha\nu\beta}+\bar g_{\mu\nu}R_{\gamma\alpha\tau\beta}-2\bar g_{\mu(\gamma}R_{\tau)\alpha\nu\beta}-2\bar g_{\nu(\gamma}R_{\tau)\alpha\mu\tau})\nonumber\\&+\dfrac{8i(\alpha-1)}{3k^6}k_{(\mu}\bar g_{\nu)(\gamma}R_{\tau)\alpha\beta\rho}k^\alpha k^\beta y^\rho+\dfrac{4(1-\alpha)}{3k^6}k_{(\mu}\bar g_{\nu)(\gamma}k_{\tau)}k^\alpha k^\beta R_{\alpha\rho\beta\sigma}y^\rho y^\sigma\nonumber\\&+\dfrac{1}{6k^4}(\bar g_{\mu\nu}\bar g_{\gamma\tau}-2\bar g_{\mu(\gamma}g_{\tau)\nu})k^\alpha k^\beta R_{\alpha\rho\beta\sigma}y^\rho y^\sigma.\label{G2g}
    \end{align}  
\section{Divergent part of $S_1^2$}\label{AppD}
After considerable calculation, we obtain the following expression for the divergences in $S_1^2$,
\begin{align}
    S_1^2|_{\text{div}}=\frac{\sqrt{|\bar g|}}{16\pi^2\varepsilon}\Bigg\{&\bar \phi^2m^4\left(\alpha-3+4\xi-3\xi^2+\frac{\xi^2}{\alpha}+3\omega-\alpha\omega-2\xi\omega-\frac{3\omega^2}{4}+\frac{\alpha\omega^2}{4}\right)+m^2\bar\phi^2 R\Big(\frac{11\xi}{3}+\alpha\xi+\frac{13\xi^2}{2}\nonumber\\&-\frac{\xi^2}{6}-6\xi^3+\frac{2\xi^3}{\alpha}+\frac{16\xi\omega}{3}-\alpha\xi\omega-10\xi^2\omega+\frac{\xi^2\omega}{\alpha}+6\xi^3\omega-\frac{2\xi^3\omega}{\alpha}-\frac{7\xi\omega^2}{4}+\frac{\alpha\xi\omega^2}{4}+2\xi^2\omega^2\Big)\nonumber\\&+m^2\bar\phi\Box\bar\phi\Big(2+\frac{1}{\alpha}-\alpha-2\frac{\xi}{\alpha}-3\xi^2+\frac{\xi^2}{\alpha}-5\omega+2\alpha\omega+2\xi\omega+\frac{3\omega^2}{2}-\frac{\alpha\omega^2}{2}\Big)\nonumber\\&+R\bar\phi\Box\bar\phi\Big(\frac{5}{6}-\frac{1}{6\alpha}-\frac{7\alpha}{6}+\frac{\alpha^2}{2}+\xi+\frac{4\xi}{3\alpha}+\frac{\xi^2}{2}-\frac{13\xi^2}{6\alpha}-3\xi^3+\frac{\xi^3}{\alpha}-\frac{5\omega}{6}+\frac{\omega}{2\alpha}+\alpha\omega-\frac{\alpha^2\omega}{6}\nonumber\\&-\frac{4\xi\omega}{3}-\frac{2\xi\omega}{\alpha}+\frac{\xi^2\omega}{2}+\frac{5\xi^2\omega}{2\alpha}+3\xi^3\omega-\frac{\xi^3\omega}{\alpha}+\frac{7\xi\omega^2}{6}-\frac{\alpha\xi\omega^2}{6}-2\xi^2\omega^2\Big)\nonumber\\&+R_{\mu\nu}\bar\phi\nabla^\mu\nabla^\nu\bar\phi\Big(3-\alpha-4\xi+\omega-2\alpha\omega+\alpha^2\omega+2\xi\omega\Big)+\bar\phi\Box^2\bar\phi\Big(1+2\omega-\alpha\omega-\frac{3\omega^2}{4}+\frac{\alpha\omega^2}{4}\Big)\Bigg\}
\end{align}
\section{Curvature expansion for derivatives acting on $\delta$-functions}\label{AppE}
The index for quantities at spacetime point $x$ are unprimed whereas for those at spacetime point $y$ are primed.
\begin{align}
    &\nabla_\alpha\delta^4(x,y)=\int \dfrac{d^4k}{(2\pi)^4}e^{iku}[ik_\alpha+\frac{i}{6}k_\alpha R_{\mu\nu}y^\mu y^\nu]\\
    &\nabla_{\alpha'}\delta^4(x,y)=\int \dfrac{d^4k}{(2\pi)^4}e^{iku}[ik_{\alpha'}+\frac{i}{6}k_\gamma(R^\gamma_{\alpha'\mu\nu}-R^\gamma_{\mu\nu\alpha'})u^\mu (u^\nu+2y^\nu)+\frac{i}{6}k_{\alpha'} R_{\mu\nu}y^\mu y^\nu]\\
    &\nabla_\alpha\nabla_\beta\delta^4(x,y) = \int \dfrac{d^4k}{(2\pi)^4}e^{iku}[-k_\alpha k_\beta +\frac{2i}{3}k_\mu R^\mu_{(\alpha\beta)\nu}x^\nu-\frac{1}{6}k_\alpha k_\beta R_{\mu\nu}y^\mu y^\nu]\\
    &\nabla_\alpha\nabla_{\beta'}\delta^4(x,y) = \int \dfrac{d^4k}{(2\pi)^4}e^{iku}[-k_\alpha k_{\beta'}+\frac{1}{6}k_\alpha k_\gamma(R^\gamma_{\mu\nu\beta'}-R^\gamma_{\beta'\mu\nu})u^\mu (u^\nu+2y^\nu)\nonumber\\&\hspace{40mm}+\frac{2i}{3}K_\gamma R^\gamma_{(\alpha\beta')\mu}x^\mu-\frac{1}{6}k_\alpha k_{\beta'}R_{\mu\nu}y^\mu y^\nu]\\
    &\nabla_{\alpha'}\nabla_{\beta'}\delta^4(x,y)=\int \dfrac{d^4k}{(2\pi)^4}e^{iku}[-k_{\alpha'}k_{\beta'}+\frac{1}{3}k_{(\alpha'}k_\gamma(R^\gamma_{\mu\nu\beta')}-R^\gamma_{\beta')\mu\nu})u^\mu (u^\nu+2y^\nu)\nonumber\\&\hspace{40mm}+\frac{2i}{3}k_\mu R^\mu_{(\alpha'\beta')\nu}x^\nu-\frac{1}{6}k_{\alpha'}k_{\beta'}R_{\mu\nu}y^\mu y^\nu]\\
    &\nabla_\alpha\nabla_\beta\nabla_\gamma\delta^4(x,y)=\int \dfrac{d^4k}{(2\pi)^4}e^{iku}[-ik_\alpha k_\beta k_\gamma+\frac{2i}{3}k_\tau R^\tau_{(\beta\gamma)\alpha}-\frac{2}{3}k_\alpha k_\tau R^\tau_{(\beta \gamma)\mu}x^\mu\nonumber\\&\hspace{40mm}-\frac{2}{3}k_\beta k_\tau R^\tau_{(\alpha \gamma)\mu}x^\mu-\frac{2}{3}k_\gamma k_\tau R^\tau_{(\beta \alpha)\mu}x^\mu-\frac{i}{6}k_\alpha k_\beta k_\gamma R_{\mu\nu}y^\mu y^\nu]\\
    &\nabla_\alpha\nabla_\beta\nabla_{\gamma'}\delta^4(x,y)=\int \dfrac{d^4k}{(2\pi)^4}e^{iku}[-ik_\alpha k_\beta k_{\gamma'}+\frac{2i}{3}k_\tau R^\tau_{(\beta\gamma')\alpha}-\frac{2}{3}k_\alpha k_\tau R^\tau_{(\beta \gamma')\mu}x^\mu\nonumber\\&\hspace{40mm}-\frac{2}{3}k_\beta k_\tau R^\tau_{(\alpha \gamma')\mu}x^\mu-\frac{2}{3}k_{\gamma'} k_\tau R^\tau_{(\beta \alpha)\mu}x^\mu-\frac{i}{6}k_\alpha k_\beta k_{\gamma'} R_{\mu\nu}y^\mu y^\nu\nonumber\\&\hspace{40mm}+\frac{i}{6}k_\alpha k_\beta k_\tau (R^\tau_{\mu\nu\gamma'}-R^\tau_{\gamma'\mu\nu})u^\mu (u^\nu+2y^\nu)]\\
     &\nabla_\alpha\nabla_{\beta'}\nabla_{\gamma'}\delta^4(x,y)=\int \dfrac{d^4k}{(2\pi)^4}e^{iku}[-ik_\alpha k_{\beta'} k_{\gamma'}+\frac{2i}{3}k_\tau R^\tau_{(\beta'\gamma')\alpha}-\frac{2}{3}k_\alpha k_\tau R^\tau_{(\beta' \gamma')\mu}x^\mu\nonumber\\&\hspace{40mm}-\frac{2}{3}k_{\beta'} k_\tau R^\tau_{(\alpha \gamma')\mu}x^\mu-\frac{2}{3}k_{\gamma'} k_\tau R^\tau_{(\beta' \alpha)\mu}x^\mu-\frac{i}{6}k_\alpha k_{\beta'} k_{\gamma'} R_{\mu\nu}y^\mu y^\nu\nonumber\\&\hspace{40mm}+\frac{i}{3}k_\alpha k_{(\beta'} k_\tau (R^\tau_{\mu\nu\gamma')}-R^\tau_{\gamma')\mu\nu})u^\mu (u^\nu+2y^\nu)]\\
      &\nabla_{\alpha'}\nabla_{\beta'}\nabla_{\gamma'}\delta^4(x,y)=\int \dfrac{d^4k}{(2\pi)^4}e^{iku}[-ik_{\alpha'} k_{\beta'} k_{\gamma'}+\frac{2i}{3}k_\tau R^\tau_{(\beta'\gamma')\alpha'}-\frac{2}{3}k_{\alpha'} k_\tau R^\tau_{(\beta' \gamma')\mu}x^\mu\nonumber\\&\hspace{40mm}-\frac{2}{3}k_{\beta'} k_\tau R^\tau_{(\alpha' \gamma')\mu}x^\mu-\frac{2}{3}k_{\gamma'} k_\tau R^\tau_{(\beta' \alpha')\mu}x^\mu-\frac{i}{6}k_{\alpha'} k_{\beta'} k_{\gamma'} R_{\mu\nu}y^\mu y^\nu\nonumber\\&\hspace{40mm}+\frac{i}{6}k_{\alpha'} k_{\beta'} k_\tau (R^\tau_{\mu\nu\gamma'}-R^\tau_{\gamma'\mu\nu})u^\mu (u^\nu+2y^\nu)\nonumber\\
      &\hspace{40mm}+\frac{i}{6}k_{\alpha'} k_{\gamma'} k_\tau (R^\tau_{\mu\nu\beta'}-R^\tau_{\beta'\mu\nu})u^\mu (u^\nu+2y^\nu)\nonumber\\&\hspace{40mm}+\frac{i}{6}k_{\beta'} k_{\gamma'} k_\tau (R^\tau_{\mu\nu\alpha'}-R^\tau_{\alpha'\mu\nu})u^\mu (u^\nu+2y^\nu)]\\
&\nabla_{\alpha}\nabla_{\beta}\nabla_{\gamma}\nabla_{\tau}\delta^4(x,y) = \int \dfrac{d^4k}{(2\pi)^4}e^{iku}[k_{\alpha}k_{\beta}k_{\gamma}k_{\tau}-\bigg\{\frac{2}{3}k_\tau k_\rho R^\rho_{(\beta\gamma)\alpha}+\frac{2}{3}k_\gamma k_\rho R^\rho_{(\beta\tau)\alpha}\nonumber\\&\hspace{40mm}+\frac{2}{3}k_\beta k_\rho R^\rho_{(\gamma\tau)\alpha}+\frac{2}{3}k_\alpha k_\rho R^\rho_{(\gamma\tau)\beta}+\frac{2i}{3}k_\alpha k_\beta k_\rho R^\rho_{(\gamma\tau)\sigma}x^\sigma\nonumber\\
&\hspace{40mm}+\frac{2i}{3}k_\alpha k_\gamma k_\rho R^\rho_{(\beta\tau)\sigma}x^\sigma+\frac{2i}{3}k_\alpha k_\tau k_\rho R^\rho_{(\beta\gamma)\sigma}x^\sigma+\frac{2i}{3}k_\beta k_\gamma k_\rho R^\rho_{(\alpha\tau)\sigma}x^\sigma\nonumber\\&\hspace{40mm}+\frac{2i}{3}k_\beta k_\tau k_\rho R^\rho_{(\alpha\gamma)\sigma}x^\sigma+\frac{2i}{3}k_\gamma k_\tau k_\rho R^\rho_{(\alpha\beta)\sigma}x^\sigma\nonumber\\
&\hspace{40mm}-\dfrac{1}{6}k_\alpha k_\beta k_\gamma k_\tau R_{\sigma\rho}y^\sigma y^\rho\bigg\}]\\
&\nabla_{\alpha}\nabla_{\beta}\nabla_{\gamma}\nabla_{\tau'}\delta^4(x,y) = \int \dfrac{d^4k}{(2\pi)^4}e^{iku}[k_{\alpha}k_{\beta}k_{\gamma}k_{\tau'}-\bigg\{\frac{2}{3}k_{\tau'} k_\rho R^\rho_{(\beta\gamma)\alpha}+\frac{2}{3}k_\gamma k_\rho R^\rho_{(\beta\tau')\alpha}\nonumber\\&\hspace{40mm}+\frac{2}{3}k_\beta k_\rho R^\rho_{(\gamma\tau')\alpha}+\frac{2}{3}k_\alpha k_\rho R^\rho_{(\gamma\tau')\beta}+\frac{2i}{3}k_\alpha k_\beta k_\rho R^\rho_{(\gamma\tau')\sigma}x^\sigma\nonumber\\
&\hspace{40mm}+\frac{2i}{3}k_\alpha k_\gamma k_\rho R^\rho_{(\beta\tau')\sigma}x^\sigma+\frac{2i}{3}k_\alpha k_{\tau'} k_\rho R^\rho_{(\beta\gamma)\sigma}x^\sigma\nonumber\\&\hspace{40mm}+\frac{2i}{3}k_\beta k_\gamma k_\rho R^\rho_{(\alpha\tau')\sigma}x^\sigma+\frac{2i}{3}k_\beta k_{\tau'} k_\rho R^\rho_{(\alpha\gamma)\sigma}x^\sigma\nonumber\\
&\hspace{40mm}+\frac{2i}{3}k_\gamma k_{\tau'} k_\rho R^\rho_{(\alpha\beta)\sigma}x^\sigma-\dfrac{1}{6}k_\alpha k_\beta k_\gamma k_{\tau'} R_{\sigma\rho}y^\sigma y^\rho\nonumber\\&\hspace{40mm}+\dfrac{1}{6}k_\alpha k_\beta k_\gamma k_\epsilon(R^\epsilon_{\sigma\rho\tau'}-R^\epsilon_{\tau'\sigma\rho})u^\sigma(u^\rho+2y^\rho)\bigg\}]\\
&\nabla_{\alpha}\nabla_{\beta}\nabla_{\gamma'}\nabla_{\tau'}\delta^4(x,y) = \int \dfrac{d^4k}{(2\pi)^4}e^{iku}[k_{\alpha}k_{\beta}k_{\gamma'}k_{\tau'}-\bigg\{\frac{2}{3}k_{\tau'} k_\rho R^\rho_{(\beta\gamma')\alpha}+\frac{2}{3}k_{\gamma'} k_\rho R^\rho_{(\beta\tau')\alpha}\nonumber\\&\hspace{40mm}+\frac{2}{3}k_\beta k_\rho R^\rho_{(\gamma'\tau')\alpha}+\frac{2}{3}k_\alpha k_\rho R^\rho_{(\gamma'\tau')\beta}+\frac{2i}{3}k_\alpha k_\beta k_\rho R^\rho_{(\gamma'\tau')\sigma}x^\sigma\nonumber\\
&\hspace{40mm}+\frac{2i}{3}k_\alpha k_{\gamma'} k_\rho R^\rho_{(\beta\tau')\sigma}x^\sigma+\frac{2i}{3}k_\alpha k_{\tau'} k_\rho R^\rho_{(\beta\gamma')\sigma}x^\sigma\nonumber\\&\hspace{40mm}+\frac{2i}{3}k_\beta k_{\gamma'} k_\rho R^\rho_{(\alpha\tau')\sigma}x^\sigma+\frac{2i}{3}k_\beta k_{\tau'} k_\rho R^\rho_{(\alpha\gamma')\sigma}x^\sigma\nonumber\\
&\hspace{40mm}+\frac{2i}{3}k_{\gamma'} k_{\tau'} k_\rho R^\rho_{(\alpha\beta)\sigma}x^\sigma-\dfrac{1}{6}k_\alpha k_\beta k_{\gamma'} k_{\tau'} R_{\sigma\rho}y^\sigma y^\rho\nonumber\\
&\hspace{40mm}+\dfrac{1}{6}k_\alpha k_\beta k_{\gamma'} k_\epsilon(R^\epsilon_{\sigma\rho\tau'}-R^\epsilon_{\tau'\sigma\rho})u^\sigma(u^\rho+2y^\rho)\nonumber\\&\hspace{40mm}+\dfrac{1}{6}k_\alpha k_\beta k_{\tau'} k_\epsilon(R^\epsilon_{\sigma\rho\gamma'}-R^\epsilon_{\gamma'\sigma\rho})u^\sigma(u^\rho+2y^\rho)\bigg\}]\\
&\nabla_{\alpha}\nabla_{\beta'}\nabla_{\gamma'}\nabla_{\tau'}\delta^4(x,y) = \int \dfrac{d^4k}{(2\pi)^4}e^{iku}[k_{\alpha}k_{\beta'}k_{\gamma'}k_{\tau'}-\bigg\{\frac{2}{3}k_{\tau'} k_\rho R^\rho_{(\beta'\gamma')\alpha}+\frac{2}{3}k_{\gamma'} k_\rho R^\rho_{(\beta'\tau')\alpha}\nonumber\\&\hspace{40mm}+\frac{2}{3}k_{\beta'} k_\rho R^\rho_{(\gamma'\tau')\alpha}+\frac{2}{3}k_\alpha k_\rho R^\rho_{(\gamma'\tau')\beta'}+\frac{2i}{3}k_\alpha k_{\beta'} k_\rho R^\rho_{(\gamma'\tau')\sigma}x^\sigma\nonumber\\
&\hspace{40mm}+\frac{2i}{3}k_\alpha k_{\gamma'} k_\rho R^\rho_{(\beta'\tau')\sigma}x^\sigma+\frac{2i}{3}k_\alpha k_{\tau'} k_\rho R^\rho_{(\beta'\gamma')\sigma}x^\sigma\nonumber\\&\hspace{40mm}+\frac{2i}{3}k_{\beta'} k_{\gamma'} k_\rho R^\rho_{(\alpha\tau')\sigma}x^\sigma+\frac{2i}{3}k_{\beta'} k_{\tau'} k_\rho R^\rho_{(\alpha\gamma')\sigma}x^\sigma\nonumber\\
&\hspace{40mm}+\frac{2i}{3}k_{\gamma'} k_{\tau'} k_\rho R^\rho_{(\alpha\beta')\sigma}x^\sigma-\dfrac{1}{6}k_\alpha k_{\beta'} k_{\gamma'} k_{\tau'} R_{\sigma\rho}y^\sigma y^\rho\nonumber\\
&\hspace{40mm}+\dfrac{1}{6}k_\alpha k_{\beta'} k_{\gamma'} k_\epsilon(R^\epsilon_{\sigma\rho\tau'}-R^\epsilon_{\tau'\sigma\rho})u^\sigma(u^\rho+2y^\rho)\nonumber\\
&\hspace{40mm}+\dfrac{1}{6}k_\alpha k_{\beta'} k_{\tau'} k_\epsilon(R^\epsilon_{\sigma\rho\gamma'}-R^\epsilon_{\gamma'\sigma\rho})u^\sigma(u^\rho+2y^\rho)\nonumber\\
&\hspace{40mm}+\dfrac{1}{6}k_\alpha k_{\gamma'} k_{\tau'} k_\epsilon(R^\epsilon_{\sigma\rho\beta'}-R^\epsilon_{\beta'\sigma\rho})u^\sigma(u^\rho+2y^\rho)\bigg\}]\\
&\nabla_{\alpha'}\nabla_{\beta'}\nabla_{\gamma'}\nabla_{\tau'}\delta^4(x,y) = \int \dfrac{d^4k}{(2\pi)^4}e^{iku}[k_{\alpha'}k_{\beta'}k_{\gamma'}k_{\tau'}-\bigg\{\frac{2}{3}k_{\tau'} k_\rho R^\rho_{(\beta'\gamma')\alpha'}+\frac{2}{3}k_{\gamma'} k_\rho R^\rho_{(\beta'\tau')\alpha'}\nonumber\\&\hspace{40mm}+\frac{2}{3}k_{\beta'} k_\rho R^\rho_{(\gamma'\tau')\alpha'}+\frac{2}{3}k_{\alpha'} k_\rho R^\rho_{(\gamma'\tau')\beta'}+\frac{2i}{3}k_{\alpha'} k_{\beta'} k_\rho R^\rho_{(\gamma'\tau')\sigma}x^\sigma\nonumber\\
&\hspace{40mm}+\frac{2i}{3}k_{\alpha'} k_{\gamma'} k_\rho R^\rho_{(\beta'\tau')\sigma}x^\sigma+\frac{2i}{3}k_{\alpha'} k_{\tau'} k_\rho R^\rho_{(\beta'\gamma')\sigma}x^\sigma\nonumber\\&\hspace{40mm}+\frac{2i}{3}k_{\beta'} k_{\gamma'} k_\rho R^\rho_{(\alpha'\tau')\sigma}x^\sigma+\frac{2i}{3}k_{\beta'} k_{\tau'} k_\rho R^\rho_{(\alpha'\gamma')\sigma}x^\sigma\nonumber\\
&\hspace{40mm}+\frac{2i}{3}k_{\gamma'} k_{\tau'} k_\rho R^\rho_{(\alpha'\beta')\sigma}x^\sigma-\dfrac{1}{6}k_{\alpha'} k_{\beta'} k_{\gamma'} k_{\tau'} R_{\sigma\rho}y^\sigma y^\rho\nonumber\\
&\hspace{40mm}+\dfrac{1}{6}k_{\alpha'} k_{\beta'} k_{\gamma'} k_\epsilon(R^\epsilon_{\sigma\rho\tau'}-R^\epsilon_{\tau'\sigma\rho})u^\sigma(u^\rho+2y^\rho)\nonumber\\
&\hspace{40mm}+\dfrac{1}{6}k_{\alpha'} k_{\beta'} k_{\tau'} k_\epsilon(R^\epsilon_{\sigma\rho\gamma'}-R^\epsilon_{\gamma'\sigma\rho})u^\sigma(u^\rho+2y^\rho)\nonumber\\
&\hspace{40mm}+\dfrac{1}{6}k_{\alpha'} k_{\gamma'} k_{\tau'} k_\epsilon(R^\epsilon_{\sigma\rho\beta'}-R^\epsilon_{\beta'\sigma\rho})u^\sigma(u^\rho+2y^\rho)\nonumber\\
&\hspace{40mm}+\dfrac{1}{6}k_{\beta'} k_{\gamma'} k_{\tau'} k_\epsilon(R^\epsilon_{\sigma\rho\alpha'}-R^\epsilon_{\alpha'\sigma\rho})u^\sigma(u^\rho+2y^\rho)\bigg\}]
\end{align}
    		

\begin{thebibliography}{0}%
\makeatletter
\providecommand \@ifxundefined [1]{%
 \@ifx{#1\undefined}
}%
\providecommand \@ifnum [1]{%
 \ifnum #1\expandafter \@firstoftwo
 \else \expandafter \@secondoftwo
 \fi
}%
\providecommand \@ifx [1]{%
 \ifx #1\expandafter \@firstoftwo
 \else \expandafter \@secondoftwo
 \fi
}%
\providecommand \natexlab [1]{#1}%
\providecommand \enquote  [1]{``#1''}%
\providecommand \bibnamefont  [1]{#1}%
\providecommand \bibfnamefont [1]{#1}%
\providecommand \citenamefont [1]{#1}%
\providecommand \href@noop [0]{\@secondoftwo}%
\providecommand \href [0]{\begingroup \@sanitize@url \@href}%
\providecommand \@href[1]{\@@startlink{#1}\@@href}%
\providecommand \@@href[1]{\endgroup#1\@@endlink}%
\providecommand \@sanitize@url [0]{\catcode `\\12\catcode `\$12\catcode
  `\&12\catcode `\#12\catcode `\^12\catcode `\_12\catcode `\%12\relax}%
\providecommand \@@startlink[1]{}%
\providecommand \@@endlink[0]{}%
\providecommand \url  [0]{\begingroup\@sanitize@url \@url }%
\providecommand \@url [1]{\endgroup\@href {#1}{\urlprefix }}%
\providecommand \urlprefix  [0]{URL }%
\providecommand \Eprint [0]{\href }%
\providecommand \doibase [0]{http://dx.doi.org/}%
\providecommand \selectlanguage [0]{\@gobble}%
\providecommand \bibinfo  [0]{\@secondoftwo}%
\providecommand \bibfield  [0]{\@secondoftwo}%
\providecommand \translation [1]{[#1]}%
\providecommand \BibitemOpen [0]{}%
\providecommand \bibitemStop [0]{}%
\providecommand \bibitemNoStop [0]{.\EOS\space}%
\providecommand \EOS [0]{\spacefactor3000\relax}%
\providecommand \BibitemShut  [1]{\csname bibitem#1\endcsname}%
\let\auto@bib@innerbib\@empty
\end{thebibliography}%


\begin{thebibliography}{99}
      \bibitem{robinson}
      S.P. Robinson \& F. Wilczek, \emph{Gravitational Correction to Running of Gauge Couplings.} \href{https://link.aps.org/doi/10.1103/PhysRevLett.96.231601}{\emph{Phys. Rev. Lett. {\bf 96}, 231601}} (2006).

    \bibitem{artur}
    A. R. Pietrykowski, \emph{Gauge Dependence of Gravitational Correction to Running of Gauge Couplings.} \href{https://link.aps.org/doi/10.1103/PhysRevLett.98.061801}{\emph{Phys. Rev. Lett. {\bf 98}, 061801}} (2007).

    \bibitem{kaluza1}
    T. Appelquist \& A. Chodos, \emph{Quantum effects in Kaluza-Klein Theories.} \href{https://link.aps.org/doi/10.1103/PhysRevLett.50.141}{\emph{Phys. Rev. Lett. {\bf 50}, 141}}. (1982)

    \bibitem{kaluza2}
      G. Kunstatter \& H.P. Leivo, \emph{On the gauge-dependence of the one-loop effective potential in self-consistent dimensional reduction.} \href{https://doi.org/10.1016/0370-2693(86)90808-7}{\emph{ Phys. Lett. B, Volume 166, Issue 3, Pages 321-324}} (1986)

    \bibitem{kaluza3}
    G. Kunstatter \& H.P. Leivo, \emph{Gauge-dependence of self-consistent dimensional reduction.} \href{https://doi.org/10.1016/0550-3213(87)90014-9}{\emph{Nucl. Phys. B, Volume 279, Issue 3-4, Pages 641-658}} (1987)

    \bibitem{gd1}
        I. Antoniadis, J. Iliopoulos \& T.N. Tomaras, \emph{Gauge invariance in quantum gravity.} \href{https://doi.org/10.1016/0550-3213(86)90402-5}{\emph{Nucl. Phys. B, Volume 267, Issue 2, Pages 497-508}} (1986)

\bibitem{gd2}
        B. de Wit \& N.D. Hari Dass, \emph{Gauge independence in quantum gravity.} \href{https://doi.org/10.1016/0550-3213(92)90478-T}{\emph{Nucl. Phys. B, Volume 374, Issue 1, Pages 99-122}} (1992)

\bibitem{unique}
	    G. A. Vilkovisky, \emph{The unique effective action in quantum field theory}, \href{https://www.sciencedirect.com/science/article/abs/pii/0550321384902281}{\emph{Nucl. Phys. B, Volume 234, Issue 1, Pages 125-137}} (1984).
		
        \bibitem{tomsbook}
	    L. E. Parker and D. J. Toms, \href{https://www.cambridge.org/in/academic/subjects/physics/theoretical-physics-and-mathematical-physics/quantum-field-theory-curved-spacetime-quantized-fields-and-gravity?format=HB&isbn=9780521877879}{\emph{Quantum Field Theory in Curved Spacetime: Quantized Fields and Gravity}}, Cambridge Monographs on Mathematical Physics, Cambridge University Press (2009).
	 
        \bibitem{Odintsovbook} 
        I. L. Buchbinder, S. D. Odintsov, and I. L. Shapiro, \href{https://www.routledge.com/Effective-Action-in-Quantum-Gravity/Buchbinder-Odintsov-Shapiro/p/book/9780750301220}{\emph{Effective Action in Quantum Gravity}}, CRC Press, Bristol, UK (1992).

    \bibitem{ourpaper1}
        S. Aashish, S. Panda, A. Tinwala \& A. Vidyarthi. \emph{Covariant Effective Action for Scalar-Tensor Theories of Gravity.} \href{https://doi.org/10.1088/1475-7516/2021/10/006}{\emph{JCAP10(2021)006}} (2021).

    \bibitem{mackaytoms}
        P. Mackay \& D. J. Toms, \emph{Quantum gravity and scalar fields.} \href{https://doi.org/10.1016/j.physletb.2009.12.032}{\emph{Phys. Lett. B, Volume 684. Issue 4-5, Pages 251-255}} (2009).

\bibitem{bunch}
        T. S. Bunch \& L. Parker, \emph{Feynman Propagator in Curved Space-Time: A Momentum Space Representation.} \href{https://doi.org/10.1103/PhysRevD.20.2499}{\emph{Phys. Rev. D {\bf 20}, 2499-2510}} (1979).

        \bibitem{bunch2}
        T.S. Bunch, \emph{Local momentum space and two-loop renormalizability of $\lambda\bar\phi^4$ field theory in curved space-time.} \href{https://doi.org/10.1007/BF00759414}{\emph{Gen Relat Gravit {\bf 13}, 711–723}} (1981).
    
   \bibitem{curvasymp}
        E. Calzetta, I. Jack \& L. Parker, \emph{Quantum gauge fields at high curvature.} \href{https://doi.org/10.1103/PhysRevD.33.953}{\emph{Phys. Rev. D, {\bf 33}, 953-977}} (1986).
    
    \bibitem{kaluza4}
        M. Awada \& D. J. Toms, \emph{Induced gravitational and gauge field actions from quantized matter fields in non-abelian Kaluza-Klein theory.}   \href{https://doi.org/10.1016/0550-3213(84)90428-0}{\emph{Nucl. Phys. B, Volume 245, Pages 161-188}} (1984).

        \bibitem{kaluza5}
        S.R. Huggins, G. Kunstatter, H. P. Leivo \& D. J. Toms, \emph{The Vilkovisky-DeWitt effective action for quantum gravity} \href{https://doi.org/10.1016/0550-3213(88)90280-5}{\emph{Nucl. Phys. B, Volume 301, Issue 4, Pages 627-660}} (1988).

        \bibitem{kaluza6}
        D. J. Toms, \emph{Induced Einstein-Maxwell action in Kaluza-Klein theory.} \href{https://doi.org/10.1016/0370-2693(83)90722-0}{\emph{Phys. Lett. B, Volume 129, Issues 1-2, Pages 31-35}} (1983).

        \bibitem{otherapp1}
        E. Calzetta, H. Salehi, \& Bei-Lok Hu, \emph{Quantum kinetic field theory in curved spacetime: Covariant Wigner function and Liouville-Vlasov equations.} \href{https://doi.org/10.1103/PhysRevD.37.2901}{\emph{Phys. Rev. D, {\bf 37}, Pages 2901-2919}} (1988).

        \bibitem{otherapp2}
        Bei-Lok Hu \& D. O'Connor, \emph{Effective Lagrangian for 4 theory in curved spacetime with varying background fields: Quasilocal approximation.}   \href{https://doi.org/10.1103/PhysRevD.30.743}{\emph{Phys. Rev. D. {\bf 30}, 743}} (1984). 

        \bibitem{otherapp3}
        Bei-Lok Hu, R. Critchley, \& A. Stylianopoulos, \emph{Finite-temperature quantum field theory in curved spacetime: Quasilocal effective Lagrangians.} \href{https://doi.org/10.1103/PhysRevD.35.510}{\emph{Phys. Rev. D {\bf 35}, 510-527}} (1987). 

        \bibitem{otherapp4}
        I. Moss, D. J. Toms, \& A. Wright, \emph{Effective action at finite temperature.} \href{https://doi.org/10.1103/PhysRevD.46.1671}{\emph{Phys. Rev. D. {\bf 46}, 1671-1679}} (1992).

        \bibitem{tomsapp1}
        D. J. Toms, \emph{Quantum gravitational contributions to quantum electrodynamics.} \href{https://doi.org/10.1038/nature09506}{\emph{Nature 468, 56–59}} (2010).

        \bibitem{tomsapp2}
        D. J. Toms, \emph{Quadratic divergences and quantum gravitational contributions to gauge coupling constants.} \href{https://doi.org/10.1103/PhysRevD.84.084016}{\emph{Phys. Rev. D. {\bf 84}}} (2011). 

        \bibitem{tomsapp4}
        D. J. Toms, \emph{Gauged Yukawa model in curved spacetime.} \href{https://doi.org/10.1103/PhysRevD.98.025015}{\emph{Phys. Rev. D {\bf 98}, 025015}} (2018).

        \bibitem{tomsapp3}
        I. Moss \& D. J. Toms, \emph{Invariants of the heat equation for non-minimal operators.} \href{https://doi.org/10.1088/1751-8113/47/21/215401}{\emph{Journal of Physics A: Mathematical and Theoretical. 47}} (2013). 

	    \bibitem{dewitt}
	    B. S. DeWitt, \emph{Quantum Field Theory and Quantum Statistics vol 1, ed I. A. Batalin, C. J. Isham and G. A. Vilkovisky} (Bristol: Hilger) p 191 (1987).

     	\bibitem{scholar}
        A. O. Barvinsky, \emph{Heat kernel expansion in the background field formalism}, \href{http://www.scholarpedia.org/article/Heat_kernel_expansion_in_the_background_field_formalism#Eq-8}{\emph{Scholarpedia}, {\bf 10}(6):31644} (2015).
    
    	 \bibitem{nonminimal}
	    A.O. Barvinsky \& G.A. Vilkovisky \emph{The generalized Schwinger-Dewitt technique in gauge theories and quantum gravity} \href{https://doi.org/10.1016/0370-1573(85)90148-6}{Physics Reports,
        Volume 119, Issue 1, Pages 1-74} (1985).
	 

        \bibitem{od1}
        I.L. Buchbinder \& S.D. Odintsov \emph{Unique effective action in Kaluza-Klein theories and spontaneous compactification} \href{https://inspirehep.net/literature/273565}{Sov. J. Nucl. Phys. 47, 377-379, Yad.Fiz. 47, 598-601} (1988).
    
        \bibitem{od2}
        I.L. Buchbinder, E.N. Kirillova \& S.D. Odintsov, \emph{The Vilkovisky effective action in the even-dimensional quantum gravity} \href{https://doi.org/10.1142/S0217732389000769}{Mod. Phys. Lett. A 4, 633-644} (1989).
    
        \bibitem{od3}
        P.M. Lavrov, S.D. Odintsov \& I.V. Tyutin, \emph{On the Unique Effective Action in Field Theory} \href{https://doi.org/10.1142/S0217732388001525}{Mod. Phys. Lett. A 3, 1273-1276} (1988).
    
        \bibitem{od4}
        I.L. Buchbinder \& S.D. Odintsov, \emph{Parametrization and gauge invariant effective action for constituent fields} \href{https://doi.org/10.1016/0370-2693(89)90533-9}{Phys. Lett. B, Volume 228, Issue 1, Pages 104-110} (1989).
    
        \bibitem{od5}
        S.D. Odintsov, \emph{The Vilkovisky Effective Action in Quantum Gravity with SU(5) Grand Unification Theory} \href{https://doi.org/10.1209/0295-5075/10/4/001}{Europhys. Lett. 10, 287-292} (1989).
    
        \bibitem{od6}
        S.D. Odintsov \& I.N. Shevchenko \emph{Unique effective action in two-dimensional induced quantum gravity} \href{https://doi.org/10.1007/BF01555530}{ Z. Phys. C 56 (1992) 315-318, Sov. Phys. J. 34 (1991) 624-627, Izv.Vuz.Fiz. 7 (1991) 74-76}.
    
        \bibitem{ourpaper2}
        S. Panda, A. Tinwala \& A. Vidyarthi,
        \emph{Covariant effective action for generalized Proca theories.} \href{https://doi.org/10.1088/1475-7516/2022/01/062}{\emph{JCAP01(2022)062}} (2022).


        \bibitem{sandeep}
        S. Aashish \& S. Panda, \emph{Covariant quantum corrections to a scalar field model inspired by nonminimal natural inflation.} \href{https://doi.org/10.1088/1475-7516/2020/06/009}{\emph{JCAP06(2020)009}} (2020).

        \bibitem{shapiro}
        B. L. Giacchini, T. Netto, and I. L. Shapiro, \emph{Vilkovisky unique effective action in quantum gravity.}\href{https://doi.org/10.1103/PhysRevD.102.106006}{\emph{Phys. Rev. D {\bf 102}, 106006}} (2020).
        
        \bibitem{fredkin}
        E. S. Fradkin \& A. A. Tseytlin, \emph{On the New Definition of Off-shell Effective Action.} \href{https://doi.org/10.1016/0550-3213(84)90075-0}{\emph{Nucl. Phys. B, Volume 234, Issue 2, Pages 509-523}} (1984). 

        \bibitem{odintsovgprop}
        I. L. Bukhbinder, S.D. Odintsov \& I. L. Shapiro, \emph{Local momentum-space representation of graviton propagators in an external gravitational field and one-loop counterterms in quantum gravity.} \href{https://doi.org/10.1007/BF00893711}{\emph{Soviet Physics Journal 27, 298–300}}  (1984).
        
        \bibitem{toms1}
        D. J. Toms, \emph{Quantization of the minimal and non-minimal vector field in curved space.} \href{https://doi.org/10.48550/arXiv.1509.05989}{\emph{arXiv.1509.05989}} (2015).

        \bibitem{hamidew}
    G. W. Gibbons, \href{https://ui.adsabs.harvard.edu/abs/2010grae.book.....H/abstract}{\emph{General Relativity: An Einstein Centenary Survey.}} S. W. Hawking \& W. Israel. Cambridge University Press (2010).

    \bibitem{gilkey}
    P. B. Gilkey, \href{https://www.routledge.com/Invariance-Theory-The-Heat-Equation-and-the-Atiyah-Singer-Index-Theorem/Gilkey/p/book/9780203749791}{\emph{Invariance theory, the heat equation and the Atiyah-Singer index theorem}}, CRC Press(2018).

    \bibitem{dewittdynamical}
    B. S. DeWitt, \emph{Dynamical Theory of Groups and Fields}. \href{https://doi.org/10.1119/1.1953053}{\emph{Am. J. Phys. 341209}} (1966).
    
	\bibitem{standard}
	   C. F. Steinwachs \& A. Y. Kamenshchik, \emph{One-loop divergences for gravity nonminimally coupled to a multiplet of scalar fields: Calculation in the Jordan frame. I. the main results}, \href{https://journals.aps.org/prd/abstract/10.1103/PhysRevD.84.024026}{\emph{Phys. Rev. D} {\bf 84}, 024026} (2011).

	    \bibitem{xtensor}
	    J. M. Martin-Garcia, \href{http://www.xact.es/xTensor/index.html}{\emph{xTensor: Fast abstract tensor computer algebra}} (2002).
		
	    \bibitem{xpert}
	    D. Brizuela, J. M. Martin-Garcia, \& G. A. M. Marugan, \emph{xPert: computer algebra for metric perturbation theory}, \href{https://link.springer.com/article/10.1007%2Fs10714-009-0773-2#citeas
		}{\emph{Gen. Relativ. Gravit.} {\bf 41}, 2415} (2009).
		
            \end{thebibliography}
\end{document}